\documentclass[twocolumn,nolinenumbers,trackchanges]{aastex701}
\usepackage{amsmath}
\usepackage{amssymb}
\usepackage{amsthm}
\usepackage{natbib,aasdefs,url,bm}
\usepackage{array}
\usepackage{float}
\usepackage{graphicx}
\usepackage{subfigure}
\usepackage[dvipsnames]{xcolor}
\hypersetup{linkcolor=red,citecolor=NavyBlue,filecolor=cyan,urlcolor=purple}

\def\be{\begin{equation}}
\def\ee{\end{equation}}
\def\ba{\begin{eqnarray}}
\def\ea{\end{eqnarray}}

\def\aap{\ {A\&A}\ }

\def\apjl{\ {ApJL}\ }
\def\apjs{\ {ApJS}\ }

\begin{document}
\title{High-energy neutrinos from shocked circumnuclear material around optically-bright and infrared-only tidal disruption events}
\shorttitle{High-energy neutrinos from TDE outflow-CNM interactions}
\shortauthors{Mukhopadhyay et al.}
\correspondingauthor{Mainak Mukhopadhyay}
\email{mainak@fnal.gov}
\author[0000-0002-2109-5315]{Mainak Mukhopadhyay}
\affiliation{Astrophysics Theory Department, Theory Division, Fermi National Accelerator Laboratory, Batavia, Illinois 60510, USA}
\altaffiliation{FERMILAB-PUB-26-0709-T}
\affiliation{Kavli Institute for Cosmological Physics, University of Chicago, Chicago, Illinois 60637, USA}
\affiliation{NSF-Simons AI Institute for the Sky (SkAI), 172 E. Chestnut St., Chicago, Illinois 60611, USA}
\email{mainak@fnal.gov}
\author[0000-0003-2579-7266]{Shigeo S. Kimura}
\affiliation{Frontier Research Institute for Interdisciplinary Sciences, Tohoku University, Sendai 980-8578, Japan}
\affiliation{Astronomical Institute, Graduate School of Science, Tohoku University, Sendai 980-8578, Japan}
\email{shigeo@astr.tohoku.ac.jp}
\author[0000-0002-9350-6793]{Tatsuya Matsumoto}
\affiliation{Department of Astronomy, School of Science, The University of Tokyo, Bunkyo-ku, Tokyo 113-0033, Japan}
\email{matsumoto@astron.s.u-tokyo.ac.jp}
\begin{abstract}
Tidal disruption events (TDEs) can launch sub-relativistic outflows that drive shocks into the surrounding circumnuclear material (CNM), providing a natural site for cosmic ray (CR) acceleration and high-energy neutrino production through hadronuclear ($pp$) interactions. We model this emission for optically bright and infrared (IR)-only TDEs, the latter motivated by a recently identified population of luminous IR transients with weak or absent optical counterparts, consistent with TDEs embedded in dusty, obscured nuclear environments. We semi-analytically compute the shock dynamics, CR acceleration and transport, and neutrino production, including the radiative cooling and compression in dense environments. While optical TDEs remain inefficient $pp$ neutrino sources with comparatively small neutrino yields, radiative compression in dense IR-only TDEs enhances the target density and allows the pp efficiency to approach the calorimetric regime, yielding up to $\sim 10\%$ of the observed diffuse neutrino flux at $100$ TeV, given the rate uncertainties. The long-lasting neutrino emission and low individual source yield motivate joint electromagnetically informed stacking searches. We forecast such searches for IceCube, IceCube-Gen2, KM3NeT, and HUNT, and conclude that a future $\sim 30\ {\rm km^3}$ neutrino observatory can reach $3\sigma$ sensitivity for nearby ($z \lesssim 0.4$) IR-only TDEs with multi-year search windows. Growing optical and IR TDE samples can therefore enable population-resolved searches that can test whether dense, obscured nuclear environments are efficient high-energy neutrino sources.
\end{abstract}
\section{Introduction}
\label{sec:intro}
\begin{figure*}
\centering
\includegraphics[width=0.9\textwidth]{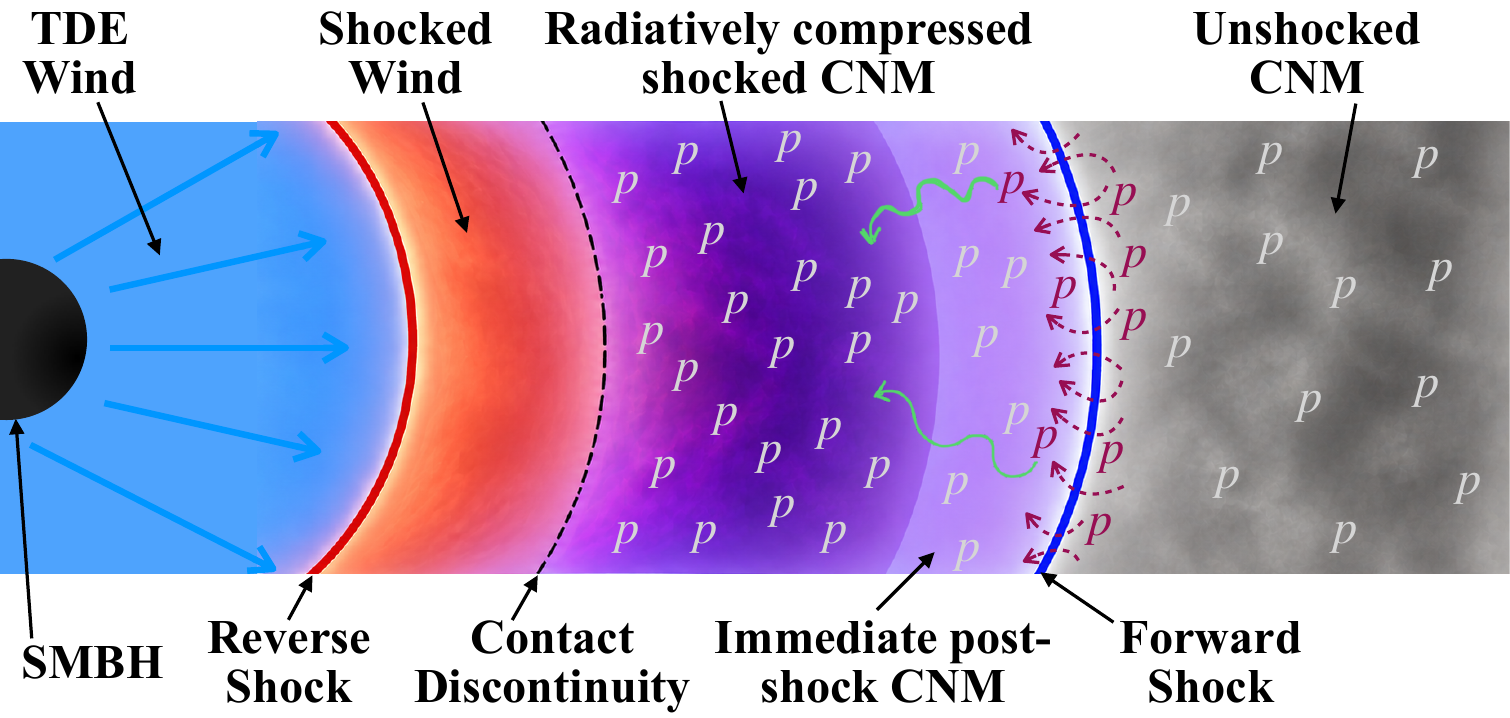}
\caption{\label{fig:schematic}Schematic illustration of the interaction between a TDE-driven outflow and the surrounding circumnuclear material (CNM). The interaction gives rise to a forward shock propagating into the unshocked CNM and a reverse shock propagating into the outflow, separated by a contact discontinuity. Cosmic-ray protons accelerated at the forward shock (purple) propagate into the shocked downstream CNM (green trajectories), where they undergo inelastic $pp$ interactions with target protons to eventually produce high-energy neutrinos. Radiative cooling can further compress the shocked CNM downstream, as illustrated by the darker region with an enhanced target proton density.
}
\end{figure*}
Tidal disruption events (TDEs) are among the most widely studied phenomena in transient astronomy. When a star passes sufficiently close to a supermassive black hole (SMBH), typically within the tidal radius where the tidal forces of the BH overcome the self-gravity of the star, the star can be disrupted. The subsequent evolution and accretion of the stellar debris can power a luminous transient known as a TDE~\citep{Rees1988,Komossa:2015qya,Stone:2018nbx}. Over the past decade, wide-field time domain surveys have rapidly increased the number of observed TDEs, particularly through optical and ultraviolet (UV) observations~\citep{Gezari:2021bmb, Yao:2023rbr}. Multiwavelength observations have also revealed TDE emission across the electromagnetic (EM) spectrum, from X-rays to radio and infrared (IR) wavelengths~\citep{Alexander:2020xwb,Gezari:2021bmb,Masterson:2024rur}. These observations probe different components and physical scales of TDEs, ranging from the innermost accretion flow to outflows and the surrounding circumnuclear environment. The TDE population is expected to grow substantially with current and upcoming wide-field surveys, providing increasingly large samples for studying their multiwavelength and high-energy properties~\citep{2019MNRAS.488.4042T,Lin:2022npv}.

A particularly interesting feature of TDEs is their ability to launch outflows. While a small subset of TDEs produce powerful relativistic jets~\citep{Burrows:2011dn,Zauderer:2011mf,DeColle:2019wzp}, non-relativistic or mildly-relativistic outflows (or winds) are inferred for a broader population from radio observations~\citep{Alexander:2020xwb,Cendes:2023rrc,Matsumoto:2021gcg,Horesh:2021gvp,Matsumoto:2024eja,Sato:2024zpq,Wu:2025vsr}. As such an outflow propagates through the circumnuclear material (CNM), it drives a forward shock into the surrounding gas. These shocks provide a natural site for the acceleration of cosmic-ray (CR) protons, which can subsequently interact with the ambient material through hadronuclear ($pp$) interactions and produce high-energy neutrinos. Unlike scenarios that rely on a powerful relativistic jet, this channel requires only a sufficiently energetic outflow and a
dense surrounding medium, making the outflow-CNM interaction a
natural target for neutrino searches from non-jetted TDEs.

The possible connection between TDEs and high-energy neutrinos has received considerable attention, motivated by a variety of neutrino-production scenarios involving relativistic or hidden jets, accretion flows and coronae, and sub-relativistic winds or outflows~\citep{Dai:2016gtz,Senno:2016bso,Lunardini:2016xwi,Hayasaki:2019kjy,Liu:2020isi,Murase:2020lnu,Winter:2020ptf,Zheng:2022afy,Mukhopadhyay:2023mld}. Several spatial and temporal coincidences between high-energy neutrinos and TDEs were previously reported~\citep{Stein:2020xhk,Reusch:2021ztx,vanVelzen:2021zsm,Winter:2022fpf,Yuan:2024foi}, although updated IceCube directional reconstructions have weakened these associations~\citep{IceCube:2025uzh,Abbasi:2026ehs}. Recent searches for neutrino emission from individual and stacked TDEs have also found no significant excess~\citep{Mukhopadhyay:2026qlq,2026arXiv260900657A}, motivating further searches using larger EM-selected TDE samples and physically motivated neutrino-emission models.

Radio observations can constrain the density structure of the CNM surrounding TDEs~\citep{Matsumoto:2021gcg,Matsumoto:2024eja}, indicating substantial diversity in their nuclear environments. Infrared searches have recently revealed a population of luminous nuclear transients with weak or absent optical counterparts, interpreted as candidate TDEs whose intrinsic optical/UV emission is absorbed and reprocessed by circumnuclear dust~\citep{Masterson:2024rur}. Their nuclear locations, characteristic IR light-curve evolution, lack of strong pre-existing AGN activity, and, in several cases, additional spectroscopic or multi-wavelength signatures support a TDE origin. The strong IR emission and absence of optical counterparts point toward efficient dust reprocessing in comparatively dusty and obscured nuclear environments, although the dust geometry and reprocessing efficiency remain uncertain.

This motivates us to consider two populations in this work: optically bright and IR-only TDEs. The denser environments of the latter are particularly interesting for high-energy neutrino production~\citep{Jiang:2023kbb,Wang:2026ijc}, since they provide enhanced target densities for hadronuclear interactions. The emergence of this potentially obscured TDE population, together with the rapidly increasing samples of optical TDEs, makes it timely to reassess the high-energy neutrino prospects of non-jetted TDEs. In particular, the combination of large EM-selected samples and next-generation neutrino observatories provides an opportunity to move beyond searches for individual source associations and systematically test the outflow–CNM scenario at the population level.

In this work, we investigate high-energy neutrino production from TDE-driven outflows interacting with the surrounding CNM. We model the forward-shock evolution and CR acceleration in optically bright and IR-only TDE environments and calculate the resulting neutrino emission from $pp$ interactions. For the dense IR-only environment, we additionally account for radiative cooling and compression of the shocked gas, which can enhance the effective target density. We calculate the neutrino emission from individual TDEs and their contribution to the diffuse astrophysical neutrino flux, and investigate EM-informed stacking searches with current and future neutrino observatories. Our results therefore connect the physical conditions inferred in TDE circumnuclear environments to their high-energy neutrino signatures and quantify the prospects for testing this scenario with the rapidly growing optical and IR-only TDE samples.

The paper is organized as follows. In Section~\ref{sec:model_shock}, we describe the interaction of TDE-driven outflows with the CNM and the resulting shock dynamics. In Section~\ref{sec:neutrino}, we calculate the high-energy neutrino emission, including the contribution to the diffuse neutrino flux. In Section~\ref{sec:search}, we investigate EM-informed neutrino search strategies. We discuss the implications and caveats of our results in Section~\ref{sec:disc} and summarize our conclusions in Section~\ref{sec:concl}.
\section{TDE outflow-CNM interaction and shock dynamics}
\label{sec:model_shock}
In this section, we describe the interaction of a TDE-driven outflow with the surrounding CNM and the resulting shock evolution. Motivated by the different environments inferred for optically-bright and IR-only TDEs, as discussed in Section~\ref{sec:intro}, we consider these two populations separately. A schematic illustration of the physical scenario is shown in Figure~\ref{fig:schematic}. The outflow launched following the tidal disruption encounters the surrounding CNM and drives a non-relativistic forward shock, where CR protons are accelerated. These accelerated protons subsequently undergo inelastic $pp$ interactions with protons in the shocked downstream material, producing pions and, ultimately, high-energy neutrinos. In the denser environments considered for IR-only TDEs, radiative cooling can further compress the shocked gas and enhance the effective target density for $pp$ interactions.

We first describe the CNM profiles adopted for the optically-bright
and IR-only TDE populations and then determine the corresponding
outflow and shock dynamics. For the dense IR-only environment, we additionally account for the transition to a radiative shock and the resulting compression of the downstream material. The resulting shock evolution and the target densities provide the physical inputs for the CR acceleration and subsequent high-energy neutrino production calculations presented in Section~\ref{sec:neutrino}.
\subsection{Properties of the circumnuclear material (CNM)}
\label{subsec:cnm}
Radio observations of TDEs provide evidence for dense circumnuclear
material surrounding the central SMBH. A natural interpretation of
the observed radio emission is synchrotron radiation produced as an
outflow launched during the TDE interacts with the surrounding CNM
and drives a non-relativistic shock. At the shock front, electrons
are accelerated and magnetic fields are amplified, giving rise to
the observed radio emission. Consequently, radio detections and
upper limits can be used to constrain both the properties of the
outflow and the density of the surrounding CNM.

The analysis of~\cite{Matsumoto:2021gcg} considered radio detections and upper limits from TDEs to constrain the local CNM density at the radius of the radio-emitting shock. The CNM density in galactic nuclei was taken to have an approximately power-law radial dependence, $n \propto r^{-k}$ with $k<3$, although a specific global density profile was not assumed in that work. Importantly, by accounting for the reduced fraction of relativistic electrons in the deep-Newtonian regime,~\cite{Matsumoto:2021gcg} inferred CNM densities considerably larger than those obtained from conventional equipartition estimates, with densities reaching $\sim 10^{5}$-$10^{6}\ {\rm cm^{-3}}$ at radii of $\sim10^{16}$-$10^{17}\ {\rm cm}$ for some radio-detected TDEs. More recently,~\cite{Matsumoto:2024eja} considered the radial structure of the surrounding medium explicitly in order to model
the late-time radio evolution of TDE outflows. In this picture, the density decreases as a power law in the inner CNM and smoothly connects to an approximately constant-density medium on scales larger than the Bondi radius.

Motivated by these results, we parameterize the CNM density as
\be
\label{eq:ncnm}
n(r) = n_{\rm ISM} \left(\frac{r}{R_B}\right)^{-k} + n_{\rm env}\,,
\ee
where $r$ is the distance from the SMBH, $k$ is the slope of the inner CNM density profile, $n_{\rm ISM}$ sets the normalization of
the inner power-law component, and $n_{\rm env}$ is the approximately constant density of the outer environment. The characteristic transition scale is taken to be the Bondi radius,
$R_B=(GM_{\rm SMBH})/c_s^2$, where $M_{\rm SMBH}$ is the SMBH mass and $c_s$ is the sound speed of the ambient nuclear gas. For an ideal gas,
$c_s=[\gamma k_{\rm B}T/(\mu m_p)]^{1/2}$, where $\gamma=5/3$ is the adiabatic index, $k_{\rm B}$ is the Boltzmann constant, $T$ is the gas temperature, $\mu=0.6$ is the mean molecular weight, and $m_p$ is the proton mass. For $M_{\rm SMBH}=10^6\,M_\odot$ and $T=10^7\ {\rm K}$, we have
$R_B \simeq 5.8\times10^{16}\ {\rm cm}$.

Similar to~\cite{Matsumoto:2024eja}, we adopt $n_{\rm ISM}=100\ {\rm cm^{-3}}$, $R_B=10^{17}\ {\rm cm}$, and $k=2.5$. For the optical TDEs, we take $n_{\rm env}=100\ {\rm cm^{-3}}$. For the IR-only TDE population, we instead consider a much denser outer environment, increasing $n_{\rm env}$ by a factor of $10^4$ relative to the optical TDEs. With the phenomenological parameterization of Equation~\eqref{eq:ncnm}, this larger $n_{\rm env}$ causes the constant-density component to dominate already at $r < R_B$. This effective transition should not be interpreted as a physical shift of the Bondi radius, but rather as a consequence of our simplified prescription for the dense IR-only environment. We emphasize that this choice is intended as an optimistic benchmark for the dense and obscured environments associated with IR-only TDEs, rather than as a uniquely determined density profile.

The motivation for considering such a dense environment comes from
the properties of IR-selected TDE candidates. In particular,~\cite{Masterson:2024rur} modeled the observed IR emission as a dust echo and inferred characteristic dust radii of $R_{\rm dust}\simeq0.05$--$0.46\ {\rm pc}$, with an average $\langle R_{\rm dust} \rangle \simeq 0.23\ {\rm pc}$. The general absence of corresponding optical flares in their high-confidence sample suggests that the intrinsic optical/UV emission can be substantially obscured by circumnuclear material. An order of magnitude estimate of the density required for such
obscuration follows from the optical depth, $\tau_{\rm opt} \simeq \kappa_{\rm opt} \rho R \simeq \kappa_{\rm opt} (m_p n) R$, where $\kappa_{\rm opt}$ is the effective optical opacity. Requiring $\tau_{\rm opt} \gtrsim 1$ gives
\be
\label{eq:nobsc}
n \gtrsim 6 \times 10^4\ {\rm cm^{-3}}\ \left(\frac{\kappa_{\rm opt}} {100\ {\rm cm^2\,g^{-1}}}\right)^{-1} \left(\frac{R}{10^{17}\ {\rm cm}}\right)^{-1}\,.
\ee
At the characteristic dust radii inferred for the IR-selected
events, $R_{\rm dust}\simeq0.05$ - $0.46\ {\rm pc}$, this simple
estimate corresponds to minimum characteristic densities of
$\sim4\times10^3$ - $4\times10^4\ {\rm cm^{-3}}$. The substantially
larger value adopted in our IR-only benchmark therefore comfortably
satisfies this order of magnitude obscuration requirement. We emphasize, however, that obscuration primarily constrains the intervening column density rather than uniquely determining the local gas density encountered by the shock. Therefore, the above estimate serves only as an order of magnitude motivation for considering a dense CNM environment. As we will illustrate in the following subsections, the high density of this environment can have important consequences for both the shock evolution and the resulting high-energy neutrino emission.
\subsection{Outflow and shock dynamics}
Following the tidal disruption of a star ($M_*=1\,M_\odot$),
approximately half of the stellar material becomes gravitationally
unbound, while the remaining bound debris can return toward the
SMBH and power an accretion-driven outflow. We therefore consider
a spherical outflow launched during the TDE with initial velocity
$v_{\rm sh,init}$ and ejecta mass $M_{\rm ej}$. As fiducial parameters, we adopt $v_{\rm sh,init}=0.1c$ and $M_{\rm ej}=0.5\ M_\odot$.

We first describe the shock evolution assuming that radiative energy losses from the shocked gas are inefficient. For simplicity, we use an energy-conserving one-zone prescription, in which the initial kinetic energy is given by
\be
E_{\rm kin}^{\rm init} = \frac{1}{2}M_{\rm ej} v_{\rm sh,init}^{2}  = 4.5 \times 10^{51}\ {\rm erg} \bigg( \frac{M_{\rm ej}}{0.5\ M_\odot} \bigg) \bigg( \frac{v_{\rm sh, init}}{0.1\ c} \bigg)^2\,,
\ee
while at later times
\be
E_{\rm kin}(t) = \frac{1}{2} \left[ M_{\rm ej}+M_{\rm sw}(R_{\rm sh}) \right] v_{\rm sh}^2\,.
\ee
Equating $E_{\rm kin}(t)$ with $E_{\rm kin}^{\rm init}$ gives
\begin{align}
\label{eq:vsh}
v_{\rm sh} &= v_{\rm sh, init} \left( \frac{M_{\rm ej}} {M_{\rm ej}+M_{\rm sw}}\right)^{1/2}\,\\
&\simeq 3 \times 10^{9}\ {\rm cm\ s^{-1}} \bigg( \frac{v_{\rm sh,init}}{0.1\ c} \bigg) \bigg( 1 + \frac{M_{\rm sw}}{M_{\rm ej}}\bigg)^{-1/2}\,\nonumber.
\end{align}
Here $M_{\rm sw}$ is the ambient mass swept up by the forward
shock as it propagates through the CNM
\be
\label{eq:msw}
M_{\rm sw}\big(R_{\rm sh}(t)\big) = 4\pi m_p \int_0^{R_{\rm sh}} dr\ r^2 n(r)\,.
\ee
For the density profile introduced in Equation~\eqref{eq:ncnm}, this becomes
\be
\label{eq:msw_analytic}
M_{\rm sw}(R_{\rm sh}) = 4\pi m_p \left[ \frac{n_{\rm ISM}R_B^kR_{\rm  sh}^{3-k}}{3-k} + \frac{n_{\rm env}R_{\rm sh}^{3}}{3} \right]\,.
\ee
The first term represents the contribution from the inner
power-law CNM, while the second is the contribution from the
approximately constant-density environment. When the power-law component dominates, $M_{\rm sw}\propto R_{\rm sh}^{3-k}$. During the initial free-expansion phase, $R_{\rm sh} \simeq v_{\rm sh}^{\rm init}t$, and hence $M_{\rm sw} \propto t^{3-k}$. For $k=2.5$, this gives $M_{\rm sw} \propto t^{1/2}$. Once the constant-density component dominates, on the other hand, $M_{\rm sw}\propto R_{\rm sh}^{3}$ and the swept-up mass grows much more rapidly with radius. For the IR-only TDE, the much larger $n_{\rm env}$ causes the constant-density component to dominate at substantially smaller radii. Consequently, the second term in Equation~\eqref{eq:msw_analytic} becomes important much earlier than in the optical TDE model.

The shock radius is evolved according to
\be
\label{eq:rsh}
R_{\rm sh}(t) = R_0 + \int_{t_{\rm min}}^t dt^\prime\ v_{\rm sh}(t^\prime)\,,
\ee
where $R_0 \simeq v_{\rm sh, init} t_{\rm min}$. At early times, when $M_{\rm sw}\ll M_{\rm ej}$, Equation~\eqref{eq:vsh} gives $v_{\rm sh}\simeq v_{\rm sh, init}$ and hence $R_{\rm sh}\simeq v_{\rm sh, init}t$. As the swept-up mass becomes comparable to the ejecta mass, the accumulated
ambient material becomes dynamically important and the shock begins to decelerate.

We begin the dynamical evolution at $t_{\min}=10^6$ s, corresponding to $R_{\rm sh} \simeq 3 \times 10^{15}$ cm for $v_{\rm sh, init} = 0.1c$. This choice restricts our calculation to radii broadly comparable to those probed by radio observations
of TDE outflows, from which the adopted inner CNM density profile is motivated. At substantially smaller radii, the structure of the
circumnuclear environment is less observationally constrained, and
we therefore do not extrapolate the density profile to arbitrarily
early times. The resulting time evolution of the ambient density encountered by the forward shock can be understood analytically. At early times, when the shock remains in approximately free expansion, $R_{\rm sh}\simeq v_{\rm sh}^{\rm init}t$. For $R_{\rm sh}\ll R_B$, the power-law component of Equation~\eqref{eq:ncnm} dominates and gives $n(R_{\rm sh}) \simeq n_{\rm ISM}(R_B/v_{\rm sh, init}t)^k\propto t^{-k}$. For our fiducial $v_{\rm sh, init} = 0.1c$ and $k = 2.5$, this becomes
\begin{align}
n(R_{\rm sh}) &\simeq 6.4\times10^{5}\ {\rm cm^{-3}}\ \left(\frac{t}{10^{6}\ {\rm s}}\right)^{-2.5}
\left(\frac{n_{\rm ISM}}{100\ {\rm cm^{-3}}}\right)
\nonumber\\
&\times \left(\frac{R_B}{10^{17}\ {\rm cm}}\right)^{2.5}
\left(\frac{v_{\rm sh, init}}{0.1c}\right)^{-2.5},
\end{align}
as long as the power-law component dominates. The characteristic time at which the shock reaches the Bondi-radius scale is
\be
t_B \simeq \frac{R_B}{v_{\rm sh}^{\rm init}} \simeq 3.3\times10^{7}\ {\rm s} \simeq 1.1\ {\rm yr}\ \left(\frac{R_B}{10^{17}\ {\rm cm}}\right) \left(\frac{v_{\rm sh}^{\rm init}}{0.1c}\right)^{-1}\,.
\ee

\begin{figure}
\centering
\includegraphics[width=0.48\textwidth]{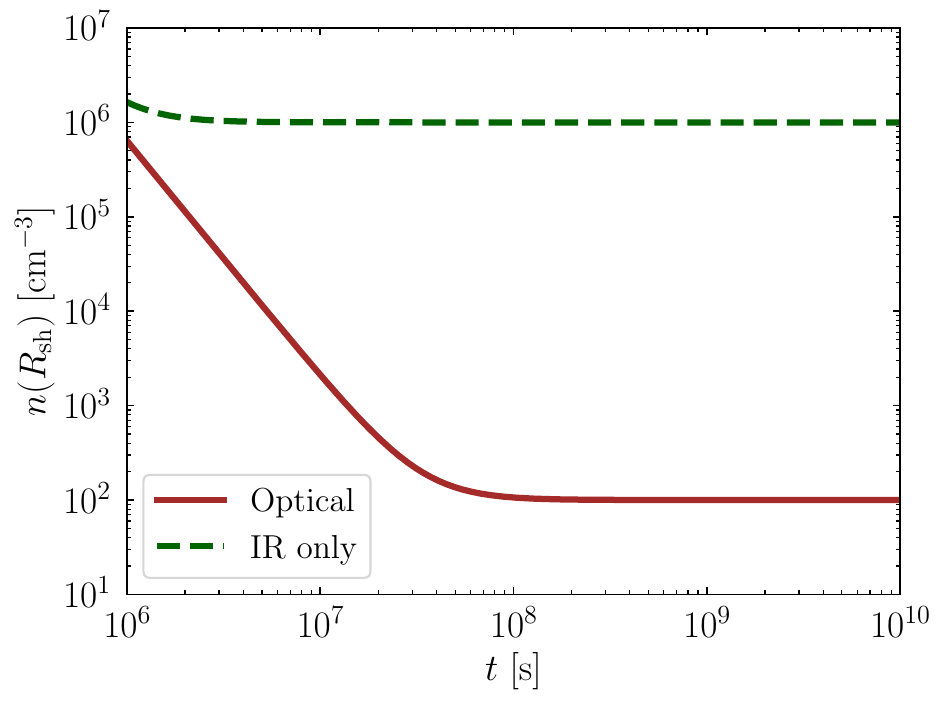}
\caption{\label{fig:density}Ambient number density encountered by the forward shock as a function of time for the optical (solid) and IR-only
(dashed) TDE models. The density follows the CNM profile given in Equation~\eqref{eq:ncnm} as the shock propagates to larger radii. At small radii, the density is dominated by the declining power-law component, while at larger radii it approaches the approximately constant environmental density, $n_{\rm env}$. The much larger $n_{\rm env}$ adopted for the IR-only population causes its density profile to flatten at substantially smaller radii and remain much denser at late times.
}
\end{figure}

Figure~\ref{fig:density} illustrates these two regimes and the substantially higher late-time density encountered by the forward shock in the IR-only TDE scenario. For the optical TDEs, $n_{\rm env}=n_{\rm ISM}$, so the two terms in Equation~\eqref{eq:ncnm} become equal at $R_{\rm sh}=R_B$ and the density subsequently approaches $n_{\rm env}=100\ {\rm cm^{-3}}$. For the IR-only model, however, the much larger $n_{\rm env}$ causes the density profile to flatten at a smaller radius, once the declining power-law component becomes comparable to the dense outer environment. The IR-only shock therefore encounters a substantially higher ambient density at early times.

\begin{figure}
\centering
\includegraphics[width=0.48\textwidth]{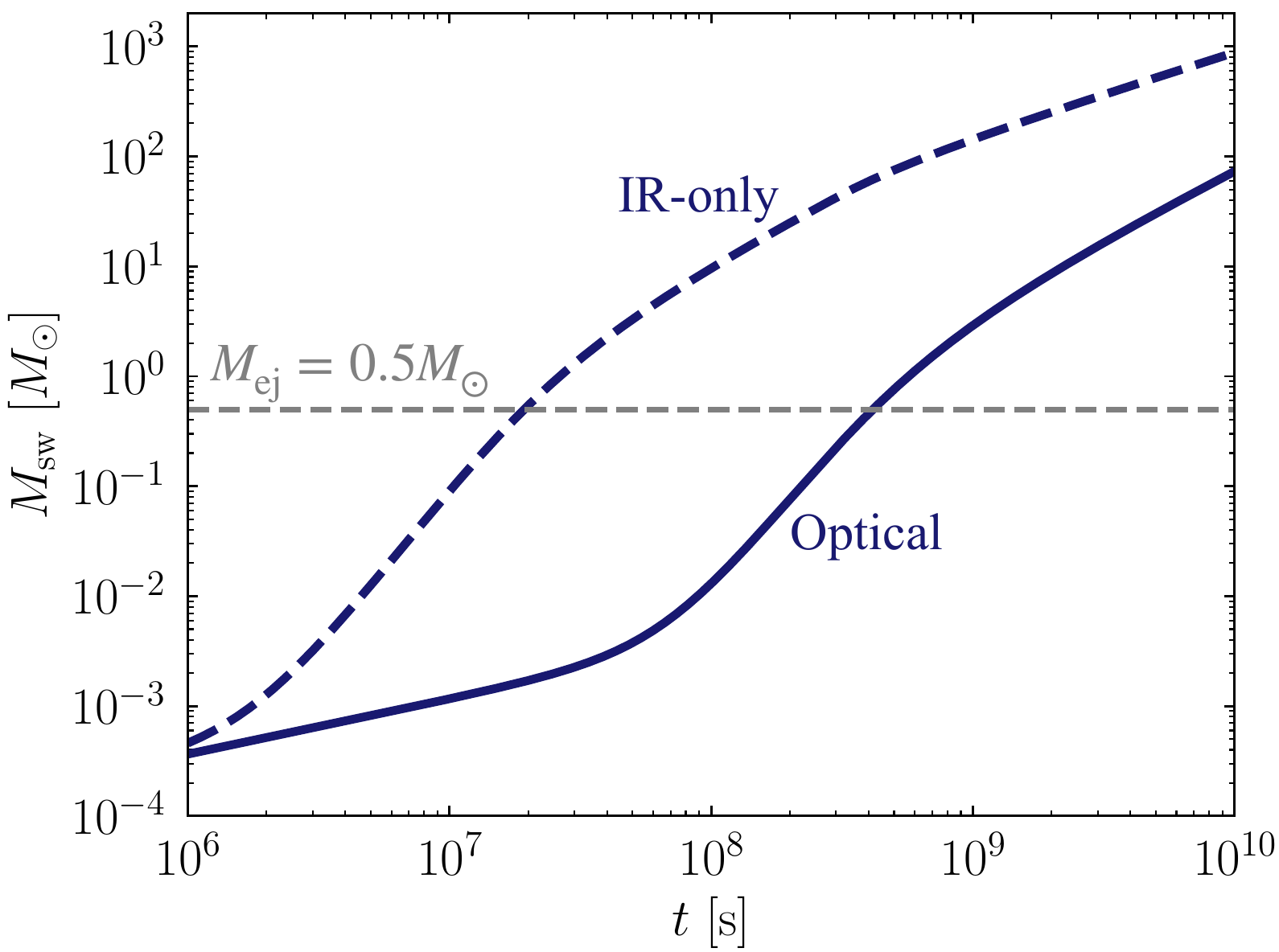}
\caption{\label{fig:mass_swept}Ambient mass swept up by the forward shock as a function of time for the optically selected (solid) and IR-only (dashed) TDE models, from Equation~\eqref{eq:msw_analytic}. The horizontal dotted line marks the fiducial ejecta mass, $M_{\rm ej} = 0.5\ M_\odot$. The larger ambient density adopted for the IR-only population causes the swept-up mass to become dynamically important substantially earlier than in the optical TDE model.
}
\end{figure}

The corresponding swept-up mass is shown in Figure~\ref{fig:mass_swept}. For the optical TDE, $n_{\rm env}=n_{\rm ISM}=100\ {\rm cm^{-3}}$, and the two density components become comparable around $R_{\rm sh}\simeq R_B$. At this radius, $M_{\rm sw}(R_B) = 4\pi m_p n_{\rm ISM}R_B^3 \big( 1/(3-k)+1/3 \big) \simeq 2.5\times10^{-3}\,M_\odot$, for $k=2.5$ and $R_B=10^{17}\ {\rm cm}$. This is much smaller than the fiducial ejecta mass $M_{\rm ej}=0.5\ M_\odot$, so the optical model remains close to free expansion as the shock crosses the Bondi-radius scale. Significant deceleration occurs only later, when $M_{\rm sw}\sim M_{\rm ej}$, at a characteristic radius of $ \sim 10^{18}\ {\rm cm}$ and a time of a few $ \times 10^8\ {\rm s}$. By contrast, Figure~\ref{fig:mass_swept} shows that $M_{\rm sw}$ approaches $M_{\rm ej}$ substantially earlier in the IR-only TDE model because of the much larger ambient mass encountered by the forward shock.

In the regime where the energy-conserving prescription is valid, the asymptotic evolution can be obtained analytically. If $M_{\rm sw}\gg M_{\rm ej}$ and the shock propagates through a constant density medium, $M_{\rm sw}\propto R_{\rm sh}^3$, so Equation~\eqref{eq:vsh} gives $v_{\rm sh}\propto R_{\rm sh}^{-3/2}$. Using $dR_{\rm sh}/dt=v_{\rm sh}$ we then have $R_{\rm sh}\propto t^{2/5}$ and $v_{\rm sh}\propto t^{-3/5}$, corresponding to the Sedov-Taylor solution~\citep{2011hea..book.....L}. Adopting the standard Sedov-Taylor normalization for an adiabatic index $\gamma = 5/3$, these scalings give
\begin{align}
R_{\rm sh} \simeq 8.8 \times 10^{16}\ {\rm cm}\  &\left(\frac{E_{\rm kin}^{\rm init}}{4.5 \times 10^{51}\ {\rm erg}}\right)^{1/5} \\
&\times \left(\frac{n_{\rm CNM}}{10^{6}\ {\rm cm^{-3}}}\right)^{-1/5} 
\left(\frac{t}{1\ {\rm yr}}\right)^{2/5}\,\nonumber,
\end{align}
and
\begin{align}
v_{\rm sh} \simeq 1.1 \times 10^{9}\ {\rm cm\ s^{-1}}\ &\left(\frac{E_{\rm kin}^{\rm init}}{4.5\times10^{51}\ {\rm erg}}\right)^{1/5} \\
&\times\left(\frac{n_{\rm CNM}}{10^{6}\ {\rm cm^{-3}}}\right)^{-1/5} 
\left(\frac{t}{1\ {\rm yr}}\right)^{-3/5}\,\nonumber,
\end{align}
We use these scalings below to illustrate the parameter dependence of the relevant physical quantities. The numerical calculations, however, employ the full shock evolution described above.

For the dense IR-only model, however, radiative cooling can become efficient during the evolution, in which case the energy-conserving
prescription in Equation~\eqref{eq:vsh} is no longer applicable. We discuss this transition and the subsequent radiative shock evolution in Section~\ref{subsec:cool_shock}.
\begin{figure}
\centering
\includegraphics[width=0.48\textwidth]{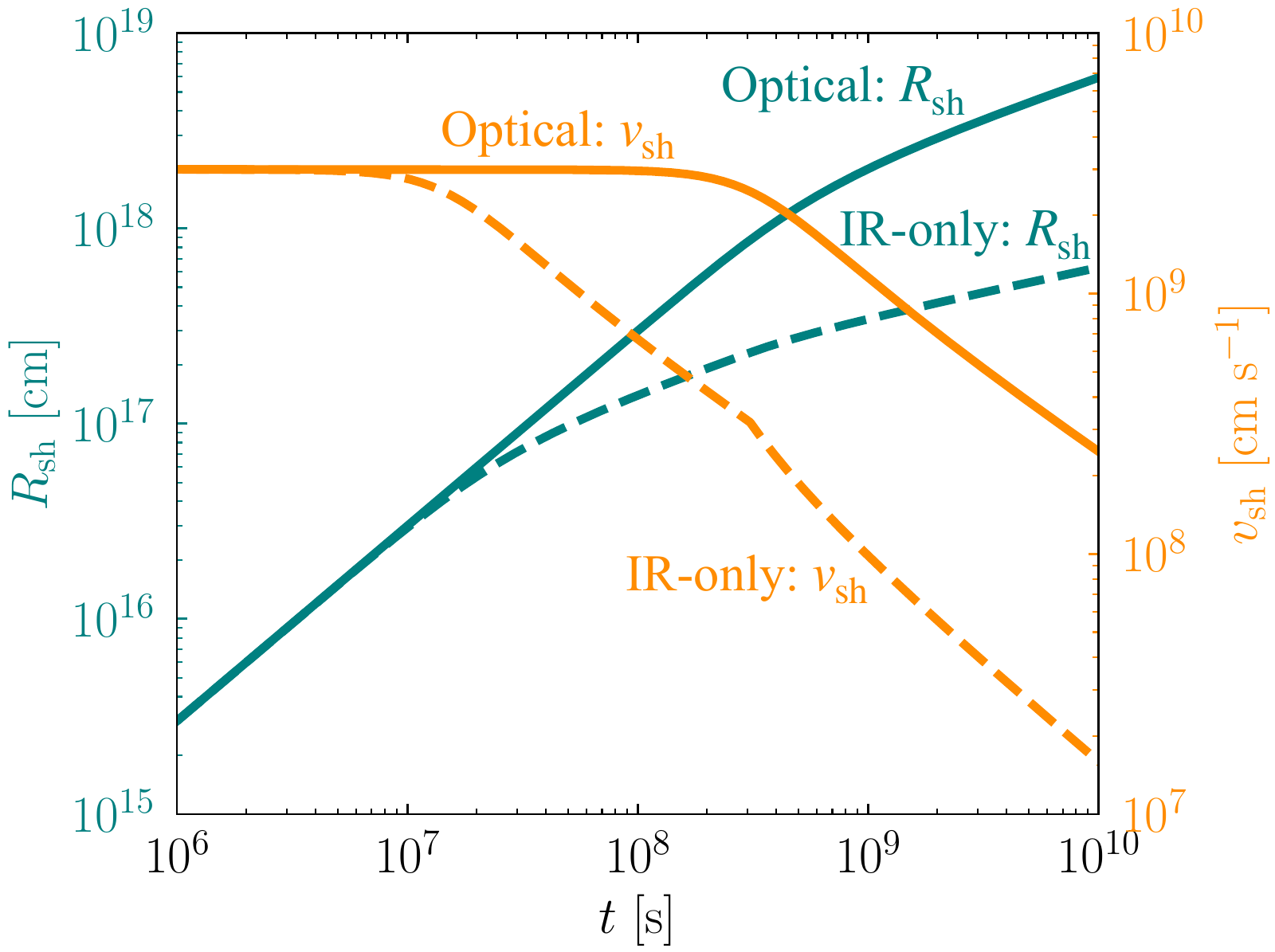}
\caption{\label{fig:rsh_vsh}Time evolution of the forward-shock radius, $R_{\rm sh}$ \emph{(left axis)}, and shock velocity, $v_{\rm sh}$ \emph{(right axis)}, for the optically selected and IR-only TDE models. The shock initially undergoes approximately free expansion, with $v_{\rm sh}\simeq v_{\rm sh}^{\rm init}$ and $R_{\rm sh}\propto t$. The larger swept-up mass in the dense IR-only environment causes the shock to decelerate substantially earlier. Once radiative cooling becomes efficient in the IR-only model, its subsequent evolution is calculated using the radiative prescription described in Section~\ref{subsec:cool_shock}.
}
\end{figure}
The numerical evolution of $R_{\rm sh}$ and $v_{\rm sh}$ is shown in Figure~\ref{fig:rsh_vsh}. The interaction of the outflow with the CNM provides the shock power available for particle acceleration as well as a dense target for hadronuclear interactions. Before calculating the
resulting high-energy neutrino emission, we first discuss the effects of radiative cooling and shock compression in the dense IR-only environment.
\subsection{Radiative cooling and shock compression in IR-only TDEs}
\label{subsec:cool_shock}
\begin{figure}
\centering
\includegraphics[width=0.48\textwidth]{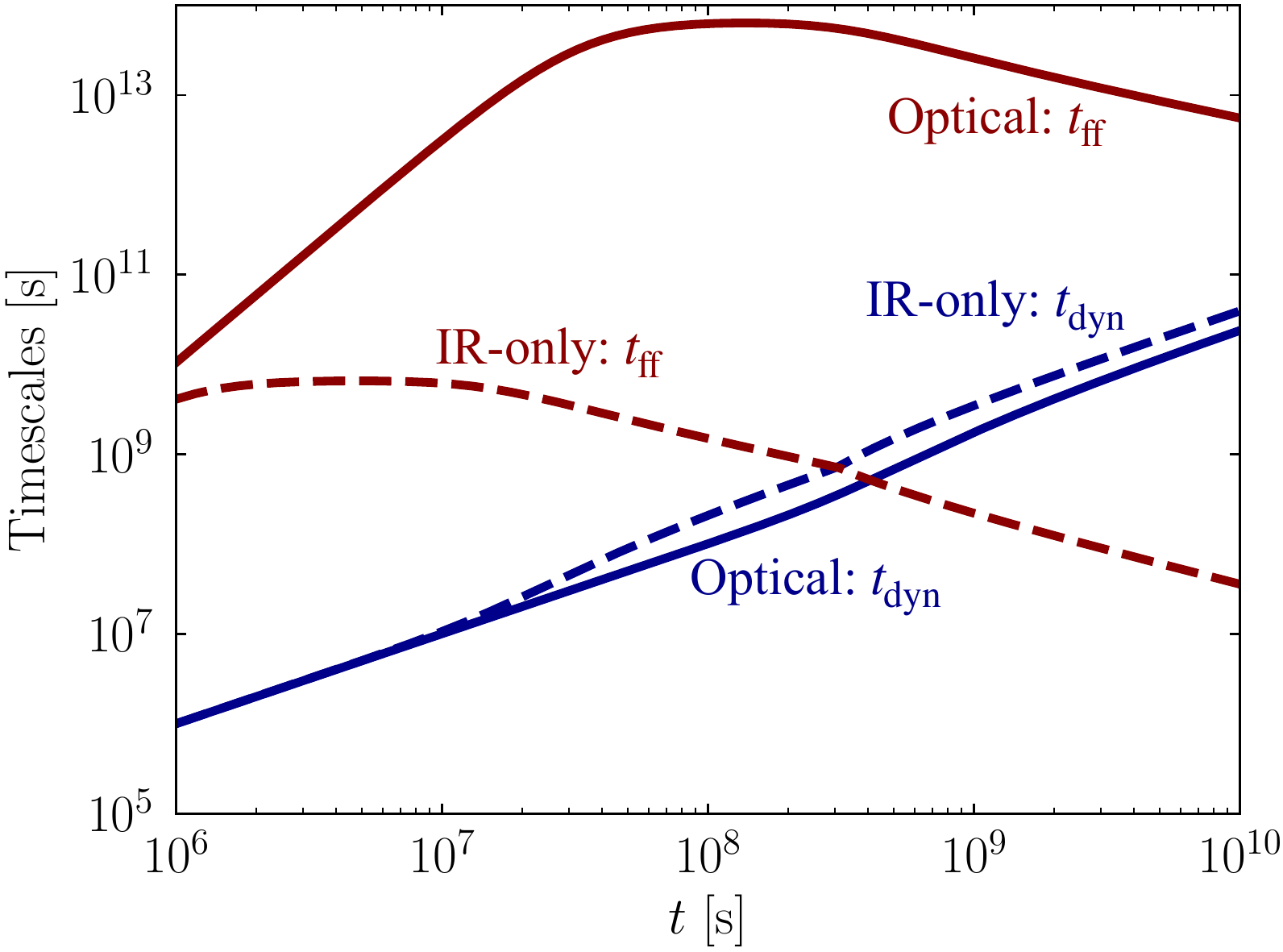}
\caption{\label{fig:tfreefree}Free-free cooling (Equation~\ref{eq:tff}) and dynamical (Equation~\ref{eq:tdyn}) timescales as functions of time for optical TDE (solid) and IR-only TDE (dashed). For the optical case, the shocked gas remains adiabatic over the timescales of interest, with $t_{\rm ff}>t_{\rm dyn}$. In the denser environment adopted for the IR-only population, free-free cooling becomes efficient once $t_{\rm ff}<t_{\rm dyn}$, resulting in the formation of a dense, radiatively cooled shell.
}
\end{figure}
In this subsection, we describe our treatment of radiative cooling
and the resulting compression of the shocked gas, which becomes
dynamically important at late times for the IR-only TDEs. The
energy-conserving shock evolution described in the previous
subsection assumes that radiative losses from the shocked gas are
inefficient. This assumption is well justified for the optical TDEs, but need not hold in the substantially denser environment adopted for the IR-only TDE population. In particular, the high post-shock density can substantially shorten the radiative cooling time, allowing the shocked gas to cool within a dynamical time. Once this occurs, the post-shock thermal energy is efficiently radiated, modifying both the subsequent shock dynamics and the density of the shocked material. We therefore account for the transition to a radiative shock in the IR-only model as described
below.

For a strong non-relativistic shock, the temperature immediately downstream of the forward shock is
\begin{align}
\label{eq:Tds}
T_{\rm ds} = \frac{3}{16} \frac{\mu m_p}{k_B} &v_{\rm sh}^{2} \simeq 1.7 \times 10^{9}\ {\rm K}\ \left(\frac{E_{\rm kin}^{\rm init}}{4.5\times10^{51}\ {\rm erg}}\right)^{2/5} \nonumber\\
&\times \left(\frac{n_{\rm CNM}}{10^{6}\ {\rm cm^{-3}}}\right)^{-2/5}
\left(\frac{t}{1\ {\rm yr}}\right)^{-6/5}\,, 
\end{align}
where $\mu=0.6$ is the mean molecular weight. For the purposes of
estimating the free-free cooling rate, we approximate the
immediate post-shock proton and electron number densities as $n_{p,\rm ds}=n_{e,\rm ds} = 4\ n_{\rm CNM}$, corresponding to the compression ratio of a strong adiabatic shock. The free-free cooling rate per unit volume is then approximated as~\citep{1986rpa..book.....R},
\be
\label{eq:Lambdaff}
\Lambda_{\rm ff} \simeq 1.7\times10^{-27} T_{\rm ds}^{1/2} n_{p,\rm ds}^2\ {\rm erg\,cm^{-3}\,s^{-1}}\,,
\ee
and the corresponding cooling time is calculated from
\begin{align}
\label{eq:tff}
t_{\rm ff} &= \frac{u_{\rm th}}{\Lambda_{\rm ff}} \approx \frac{3 n_{p,\rm ds} k_{\rm B}T_{\rm ds}}{\Lambda_{\rm ff}} \simeq 2.5 \times 10^{9}\ {\rm s}\ \left(\frac{E_{\rm kin}^{\rm init}}{4.5 \times 10^{51}\ {\rm erg}}\right)^{1/5} \nonumber \\
&\times \left(\frac{n_{\rm CNM}}{10^{6}\ {\rm cm^{-3}}}\right)^{-6/5}\left(\frac{t}{1\ {\rm yr}}\right)^{-3/5}\,,
\end{align}
where $u_{\rm th}$ is the thermal energy density in the post-shock region. With these assumptions, $t_{\rm ff}\propto T_{\rm ds}^{1/2}/n_{\rm CNM}$.
Since $T_{\rm ds}\propto v_{\rm sh}^{2}$, this gives $t_{\rm ff} \propto v_{\rm sh}/n_{\rm CNM}$. The substantially larger density of the IR-only environment therefore leads to a much shorter free-free cooling time than in the optically selected TDE model.

To determine when radiative cooling becomes dynamically important,
we compare $t_{\rm ff}$ with the dynamical time,
\be
\label{eq:tdyn}
t_{\rm dyn} = \frac{R_{\rm sh}}{v_{\rm sh}} = \frac{5}{2}t \simeq 7.9 \times 10^{7}\ {\rm s} \left(\frac{t}{1\ {\rm yr}}\right)\,,
\ee
where the second equality follows from the Sedov-Taylor regime. In the Sedov-Taylor regime, the above scalings imply $t_{\rm ff}/t_{\rm dyn} \propto (E_{\rm kin}^{\rm init})^{1/5} n_{\rm CNM}^{-6/5} t^{-8/5}$, such that radiative cooling becomes increasingly important as the shock evolves. Equating $t_{\rm ff} = t_{\rm dyn}$ gives an approximate transition time
\be
t_{\rm crit} \simeq 2.8\times 10^8\ {\rm s}\ \left(\frac{E_{\rm kin}^{\rm init}}
{4.5\times10^{51}\ {\rm erg}}\right)^{1/8} \left(\frac{n_{\rm CNM}} {10^{6}\ {\rm cm^{-3}}}\right)^{-3/4}\,.
\ee
In the numerical calculation, we determine $t_{\rm crit}$
directly from the condition $t_{\rm ff}=t_{\rm dyn}$ using the
full shock evolution.

As shown in Figure~\ref{fig:tfreefree}, the optical model satisfies $t_{\rm ff}>t_{\rm dyn}$ throughout the evolution considered here and therefore remains effectively adiabatic. For the IR-only model,
by contrast, $t_{\rm ff}$ eventually decreases below
$t_{\rm dyn}$. We define the onset of the radiative phase,
$t_{\rm crit}$, as the first time for which $t_{\rm ff}(t_{\rm crit}) = t_{\rm dyn}(t_{\rm crit})$. Prior to $t_{\rm crit}$, we use the energy-conserving shock evolution given by Equation~\eqref{eq:vsh}. At later times,
$t>t_{\rm crit}$, radiative losses become dynamically important for the shocked gas and we treat its subsequent bulk evolution as radiative. We emphasize that this classification refers to the global evolution of the shocked material, while the immediate shock transition remains well described by the adiabatic strong-shock jump conditions, with the shocked gas subsequently cooling and compressing downstream.

Once radiative cooling becomes dynamically important, the thermal
energy generated behind the shock is efficiently lost and the
energy-conserving prescription for the bulk evolution is no longer
appropriate. As a simple approximation to the subsequent evolution, we treat the shocked material as a momentum-conserving shell. We require the shell momentum to match continuously onto the pre-radiative solution at $t_{\rm crit}$, such that
\be
\label{eq:momentum_cooling}
\big( M_{\rm ej}+M_{\rm sw}(t) \big) v_{\rm sh}(t) = \big( M_{\rm ej}+M_{\rm sw}(t_{\rm crit}) \big) v_{\rm sh}(t_{\rm crit})\,.
\ee
The shock velocity during the radiative phase is therefore
\be
\label{eq:vsh_radiative}
v_{\rm sh}(t) = v_{\rm sh}(t_{\rm crit})
\bigg( \frac{M_{\rm ej}+M_{\rm sw}(t_{\rm crit})}{M_{\rm ej}+M_{\rm sw}(t)} \bigg)\,.
\ee
In the limit $M_{\rm sw} \gg M_{\rm ej}$, this approaches
$M_{\rm sw} v_{\rm sh} \simeq{\rm constant}$. More generally, for a
power-law ambient density $n_{\rm CNM}\propto R^{-k}$, such that
$M_{\rm sw}\propto R_{\rm sh}^{3-k}$, momentum conservation gives $R_{\rm sh} \propto t^{1/(4-k)}$ and $v_{\rm sh}\propto t^{-(3-k)/(4-k)}$.
The IR-only model is already dominated by its approximately
constant-density component over the radii relevant to the
radiative transition. In this limit, $k\simeq0$, and the
momentum-conserving solution approaches $R_{\rm sh}\propto t^{1/4}$ and $v_{\rm sh}\propto t^{-3/4}$, rather than the energy-conserving scalings
$R_{\rm sh}\propto t^{2/5}$ and $v_{\rm sh}\propto t^{-3/5}$ discussed in the previous subsection.

In addition to modifying the shock dynamics, radiative cooling can substantially compress the shocked material. Immediately behind the shock, the density is approximately $4n_{\rm CNM}$ according to the strong-shock jump conditions. As this material advects downstream and cools, its density can increase considerably beyond this value. We distinguish these two
regions following the physical picture discussed by~\cite{Kimura:2024lvt} (see also~\citealt{Kashiyama:2014rza}), where particle acceleration occurs around the immediate shock, whose properties are described by the adiabatic shock jump conditions, while strong density compression develops farther downstream over the radiative cooling layer.

To estimate the density of the cooled material, we assume that
the post-shock gas cools approximately isobarically to a
characteristic temperature $T_{\rm c}=10^4\ {\rm K}$ (see Section \ref{subsec:caveats} for a discussion on this treatment). Under
approximately constant thermal pressure, $nT\simeq{\rm constant}$,
so that the additional compression relative to the immediate
post-shock density is $\chi_{\rm cool} = \max \left[ 1, (T_{\rm ds}/T_{\rm c}) \right]$. We therefore approximate the number density of the radiatively
cooled shocked material as $n_{\rm cool} = 4 n_{\rm CNM} \chi_{\rm cool}$. For $T_{\rm ds} \gg T_{\rm c}$, the density of the cooled layer can
thus substantially exceed the immediate post-shock density.

It is important to distinguish this cooled-shell density from the
density relevant to particle acceleration. The large compression develops only after the shocked material has cooled downstream, whereas the physical conditions in the immediate vicinity of the forward shock remain described
by the adiabatic shock jump conditions. We therefore retain the
immediate post-shock density when evaluating particle acceleration,
while the compressed density provides the target density for CRs interacting with the radiatively cooled shocked material. In the following subsection, we discuss how this distinction modifies the hadronuclear interaction efficiency and the maximum energy of the accelerated protons.
\section{High-energy neutrino signatures}
\label{sec:neutrino}
\begin{figure}
\centering
\includegraphics[width=0.48\textwidth]{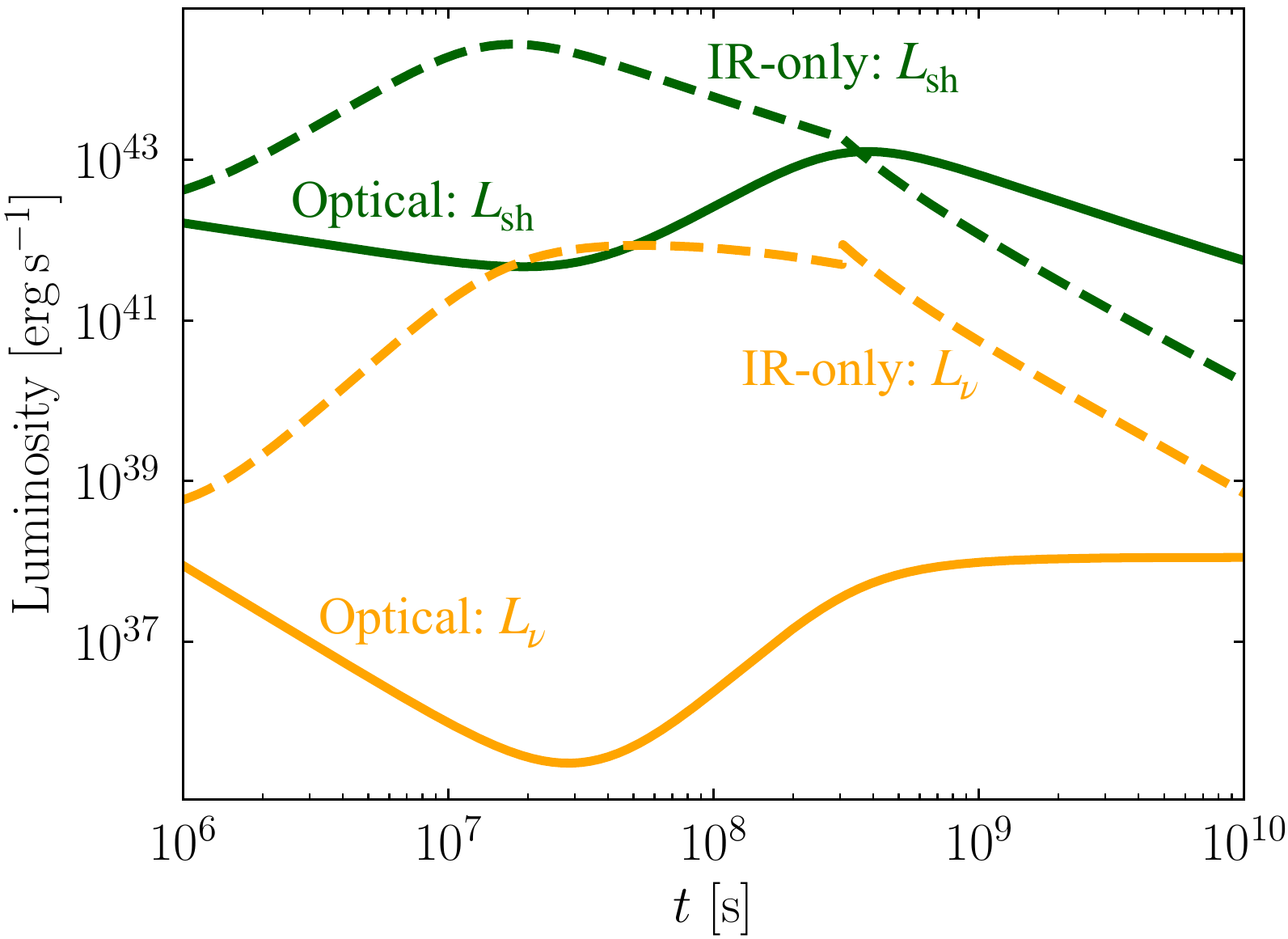}
\caption{\label{fig:luminosity}
Time evolution of the shock luminosity $L_{\rm sh}$ (Equation~\ref{eq:lsh}) and the neutrino luminosity $L_\nu$ (Equation~\ref{eq:lnu}), for the optically-bright
(solid) and IR-only (dashed) TDE models. The much denser environment of the IR-only model results in a substantially larger shock power and $pp$ interaction efficiency. At late times, radiative cooling modifies the shock evolution and the maximum energy of the accelerated protons in the IR-only model, as discussed in Section~\ref{subsec:cool_shock}.
}
\end{figure}
In this section, we calculate the high-energy neutrino emission resulting from the interaction of CR protons accelerated at the forward shock with the shocked CNM. Using the shock evolution derived in Section~\ref{sec:model_shock}, we first determine the power available for CR acceleration, the maximum proton energy, and the efficiency of hadronuclear ($pp$) interactions. The inelastic $pp$ interactions produce both neutral and charged pions, with the latter subsequently decaying into high-energy neutrinos. The substantially different CNM environments of the optically-bright and IR-only TDE populations lead to correspondingly different $pp$ interaction efficiencies. In particular, while the optical TDEs remain optically thin to $pp$ interactions over the timescales considered here, the dense shocked material in the IR-only model can approach the calorimetric regime. We then calculate the resulting neutrino spectra and time-integrated fluences and estimate the contribution of both TDE populations to the all-sky diffuse neutrino intensity. In Section~\ref{subsec:pgamma}, we additionally discuss photohadronic ($p\gamma$) interactions.
\subsection{Cosmic-ray acceleration and hadronuclear ($pp$) interactions}
\label{subsec:cr_acc}
The kinetic power processed by the forward shock, or the shock luminosity, is given by
\begin{align}
\label{eq:lsh}
L_{\rm sh} \approx \frac{1}{2}(4\pi) m_p & n_{\rm CNM} v_{\rm sh}^{3}R_{\rm sh}^{2} \simeq 1.2\times10^{44}\ {\rm erg\ s^{-1}} \nonumber\\
&\times \left(\frac{E_{\rm kin}^{\rm init}}{4.5\times10^{51}\ {\rm erg}}\right)\left(\frac{t}{1\ {\rm yr}}\right)^{-1}\,,
\end{align}
where $n_{\rm CNM}$ denotes the upstream CNM density encountered by the forward shock. We assume that a fraction $\eta_{\rm CR}=0.1$ of this power is injected into non-thermal CR protons, such that $L_{\rm CR} = \eta_{\rm CR}L_{\rm sh}$. The resulting shock luminosities are shown in Figure~\ref{fig:luminosity}. Note that, in the constant-density Sedov-Taylor limit, $L_{\rm sh} \propto E_{\rm kin}^{\rm init}\ t^{-1}$, with the explicit dependence on the ambient density canceling between the shock radius and velocity.

For the optical TDEs, the qualitative evolution of $L_{\rm sh}$ follows directly from the density and shock dynamics discussed in Section~\ref{subsec:cnm}. At early times, while the shock is approximately freely expanding through the power-law CNM, $R_{\rm sh}\propto t$, $v_{\rm sh}\simeq{\rm constant}$, and $n_{\rm CNM}\propto R_{\rm sh}^{-k}$. Thus from Equation~\eqref{eq:lsh} we have $L_{\rm sh}\propto t^{2-k}$. For $k=2.5$, $L_{\rm sh}\propto t^{-1/2}$, producing the initial decline seen in Figure~\ref{fig:luminosity}. Once the approximately constant-density component becomes important, the increasing shock surface area causes $L_{\rm sh}$ to rise. At late times, the optical TDE model remains approximately energy conserving; in the constant-density limit,
$R_{\rm sh} \propto t^{2/5}$ and $v_{\rm sh} \propto t^{-3/5}$, giving $L_{\rm sh} \propto R_{\rm sh}^{2}v_{\rm sh}^{3} \propto t^{-1}$.

The evolution of the IR-only model is different because its dense
constant-density component ($n_{\rm env}$) becomes important at much smaller radii and causes substantially earlier shock deceleration. Moreover, once $t_{\rm ff}<t_{\rm dyn}$, the shock enters the radiative regime and its subsequent dynamics are calculated using the momentum-conserving prescription of Section~\ref{subsec:cool_shock}. 

We estimate the magnetic field in the immediate post-shock
acceleration region by assuming that a fraction $\epsilon_B$ of
the shock energy density is carried by magnetic fields,
\begin{align}
\label{eq:Bfield}
B &= \sqrt{8\pi\epsilon_B n_{\rm CNM}m_pv_{\rm sh}^{2}} \simeq 0.73\ {\rm G}\ \left(\frac{\epsilon_B}{0.01}\right)^{1/2}\\
&\times \left(\frac{E_{\rm kin}^{\rm init}}{4.5\times10^{51}\ {\rm erg}}\right)^{1/5}
\left(\frac{n_{\rm CNM}}{10^{6}\ {\rm cm^{-3}}}\right)^{3/10}
\left(\frac{t}{1\ {\rm yr}}\right)^{-3/5}\,\nonumber,
\end{align}
where we adopt $\epsilon_B=0.01$. This treatment implicitly assumes that the plasma in the vicinity of the forward shock is sufficiently ionized to sustain the magnetic turbulence required for efficient particle acceleration. In a partially ionized upstream, ion-neutral damping can suppress such turbulence and thereby limit particle acceleration~\citep{Drury:1995ep,Reville:2007er,Metzger:2015zka}. We discuss this uncertainty further in Section~\ref{subsec:caveats}.

We assume diffusive shock acceleration, for which the acceleration timescale is given by $t_{\rm acc} = \big( \eta \varepsilon_p c \big)/\big(e B v_{\rm sh}^{2} \big)$, and adopt $\eta=20/3$, corresponding to the acceleration timescale for Bohm diffusion at a strong non-relativistic shock~\citep{1983RPPh...46..973D}. For reference, in the Sedov-Taylor regime this gives $t_{\rm acc} \simeq 1.5 \times 10^{6}\ {\rm s}\ \big(\varepsilon_p/2\ {\rm PeV}\big) \big(\epsilon_B/0.01 \big)^{-1/2}
\big(E_{\rm kin}^{\rm init}/(4.5\times10^{51}\ {\rm erg})\big)^{-3/5}
\big(n_{\rm CNM}/10^{6}\ {\rm cm^{-3}}\big)^{1/10}
\big(t/1\ {\rm yr} \big)^{9/5}$ for $\eta=20/3$. We account for both advective and diffusive escape from the acceleration region. Taking the advective timescale to be order $t_{\rm dyn}$ and adopting Bohm diffusion, with the diffusion coefficient given by $D(\varepsilon_p) = (1/3) c r_L = \big( c \varepsilon_p \big)/\big(3 e B \big)$, the diffusion timescale is $t_{\rm diff} = R_{\rm sh}^{2}/ \big( 6 D(\varepsilon_p) \big)$, and the total escape timescale is $t_{\rm esc}^{-1} = t_{\rm dyn}^{-1} + t_{\rm diff}^{-1}$. The proton synchrotron cooling timescale is $t_{p,\rm syn} = \big( 6 \pi m_p^{4} c^{3}\big) / \big(
\sigma_Tm_e^{2}B^{2}\varepsilon_p \big)$.

The maximum proton energy is determined by requiring the
acceleration timescale to be shorter than the relevant escape,
interaction, and cooling timescales. We determine $\varepsilon_p^{\rm max}$ by solving
\be
\label{eq:epmax}
t_{\rm acc}(\varepsilon_p^{\rm max}) = \min\left[ t_{\rm esc }(\varepsilon_p^{\rm max}), t_{pp,\rm acc}, t_{p,\rm syn}(\varepsilon_p^{\rm max}), t_{\rm ff} \right]\,,
\ee
where the $pp$ loss timescale entering the determination of
$\varepsilon_p^{\rm max}$ is evaluated using the immediate
post-shock density, since particle acceleration occurs in the vicinity of the forward shock, that is, $t_{pp,\rm acc} = 1/\left( 4 n_{\rm CNM}\sigma_{pp}\kappa_{pp}c \right)$, where we adopt the inelastic $pp$ interaction cross-section  $\sigma_{pp} = 3 \times 10^{-26}\ {\rm cm^2}$ and $\kappa_{pp}=0.5$ is the proton inelasticity. Note that since $t_{\rm esc} \leq t_{\rm dyn}$, free-free cooling cannot limit the maximum proton energy while $t_{\rm ff} > t_{\rm dyn}$, therefore it becomes a potentially relevant acceleration timescale only as the IR-only shock approaches the radiative regime. For example, if acceleration is limited by the dynamical timescale, the Sedov-Taylor scalings give
\begin{align}
\varepsilon_{p,\rm dyn}^{\rm max} &\simeq 1.1 \times 10^{8}\ {\rm GeV}\ \left(\frac{\epsilon_B}{0.01}\right)^{1/2}
\left(\frac{E_{\rm kin}^{\rm init}}{4.5\times10^{51}\ {\rm erg}}\right)^{3/5} \nonumber\\
&\times \left(\frac{n_{\rm CNM}}{10^{6}\ {\rm cm^{-3}}}\right)^{-1/10}
\left(\frac{t}{1\ {\rm yr}}\right)^{-4/5}\,.
\end{align}

The free-free cooling time in Equation~\eqref{eq:epmax} does not correspond to an energy loss channel for the relativistic protons. Rather, it characterizes the timescale over which the hot immediate post-shock gas cools and hence the duration over which the corresponding acceleration environment can be maintained. For the optical TDEs, $t_{\rm ff} > t_{\rm dyn}$ throughout the evolution considered here, and radiative cooling
does not significantly modify the shocked gas. For the IR-only TDEs, however, $t_{\rm ff}$ becomes shorter as a consequence of the much denser CNM and eventually becomes dynamically important, as shown in Figure~\ref{fig:tfreefree}. When free-free cooling
provides the shortest timescale limiting particle acceleration, $t_{\rm acc} = t_{\rm ff}$ gives $\varepsilon_{p,\rm ff}^{\rm max} = \big( eBv_{\rm sh}^{2}t_{\rm ff}\big)/(\eta c)$. Since $B \propto n_{\rm CNM}^{1/2}v_{\rm sh}$ and $t_{\rm ff} \propto v_{\rm sh}/n_{\rm CNM}$, the corresponding cooling-limited maximum energy scales as $\varepsilon_{p,\rm ff}^{\rm max} \propto v_{\rm sh}^{4}n_{\rm CNM}^{-1/2}$.
Once radiative cooling becomes dynamically important in the
IR-only TDE model, the shocked material also undergoes the strong
downstream compression discussed in Section~\ref{subsec:cool_shock}. This compression does not enter $t_{pp,\rm acc}$, since the accelerated protons are produced in the immediate shock region before the large density enhancement develops. We therefore continue to evaluate $t_{pp,\rm acc}$ using the immediate post-shock density $4n_{\rm CNM}$.

We note that this cooling-time constraint can equivalently be interpreted in terms of the finite spatial extent of the particle-acceleration region in a radiative shock. For a laminar downstream flow, the hot post-shock layer has a characteristic cooling length $\Delta_{\rm rad} \simeq v_{\rm ds}t_{\rm ff}$, where $v_{\rm ds} \simeq v_{\rm sh}/4$ for a strong shock. Requiring the downstream diffusion length $\ell_{\rm diff,ds} \simeq D/v_{\rm ds}$, to remain within this layer gives (assuming Bohm diffusion) $\varepsilon_p^{\rm max}$ that differs only by an order unity factor from that obtained using $t_{\rm acc} = t_{\rm ff}$. Thus, our treatment approximately accounts for the finite laminar cooling length once the shock becomes radiative, rather than assuming an acceleration region of size $R_{\rm sh}$. Possible modifications to this picture due to the multidimensional structure of radiative shocks are discussed in Section~\ref{subsec:caveats}.

As discussed in Section~\ref{subsec:cool_shock}, we distinguish between the immediate post-shock region, where CR acceleration takes place, and the radiatively compressed material farther downstream. For the optical TDEs, for which radiative compression is inefficient, we take the target density for $pp$ interactions to be the immediate post-shock density, $n_{\rm target}=4n_{\rm CNM}$. For the IR-only TDE, once the shock becomes radiative we instead use the density of the cooled shocked material, $n_{\rm target} = n_{\rm cool} = 4 n_{\rm CNM}\chi_{\rm cool}$ for $t > t_{\rm crit}$, where $\chi_{\rm cool}$ is the radiative compression factor defined in Section~\ref{subsec:cool_shock}.

\begin{figure}
\centering
\includegraphics[width=0.48\textwidth]{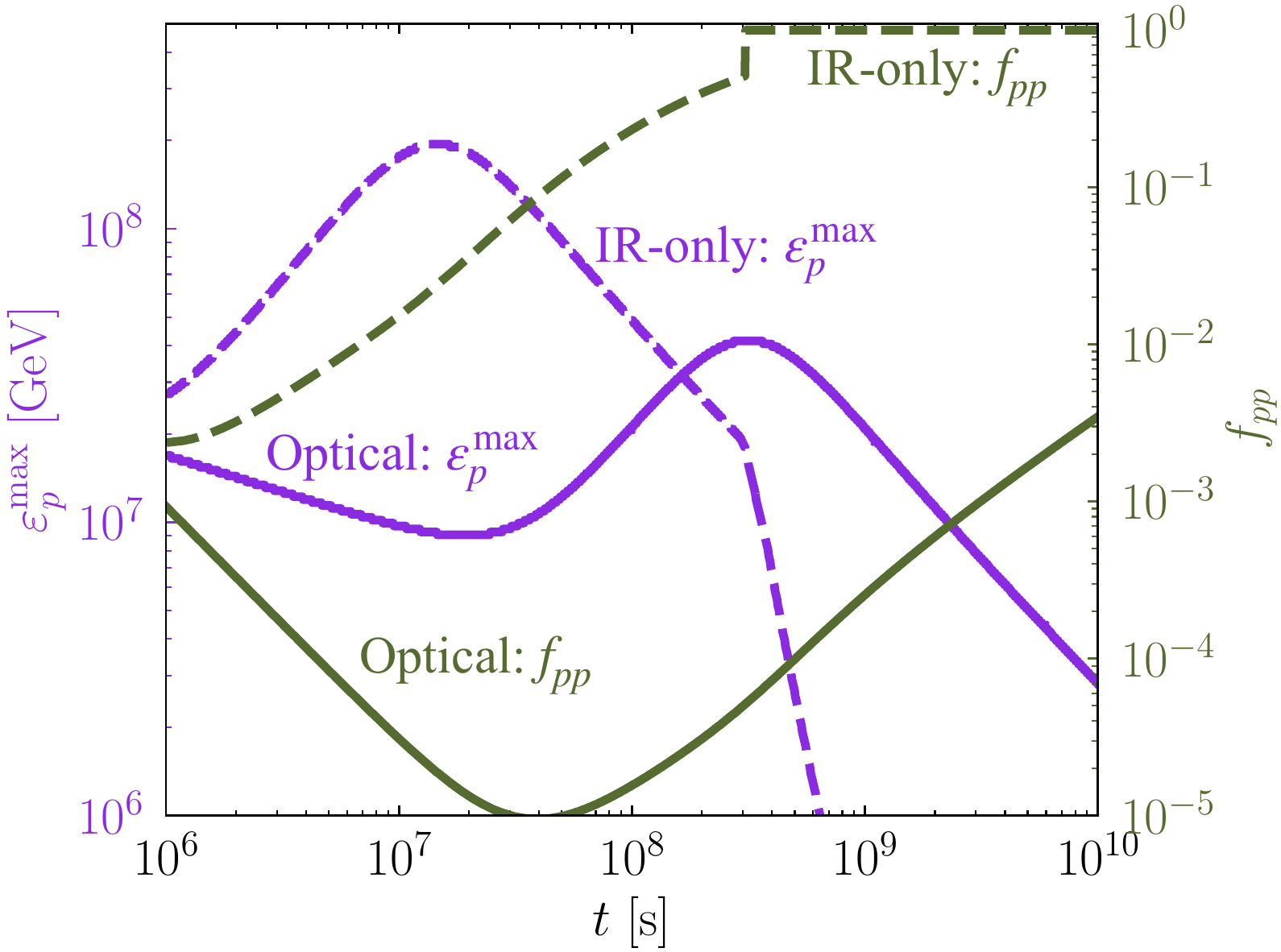}
\caption{\label{fig:epmax_fpp}Time evolution of the maximum proton energy,
$\varepsilon_p^{\rm max}$ (\emph{left axis}), and the effective
$pp$ interaction efficiency, $f_{pp}$ (\emph{right axis},
Equation~\ref{eq:fpp}), for the optically-bright (solid) and IR-only (dashed) TDE models. The $pp$ interaction efficiency is evaluated at the instantaneous maximum proton energy, $\varepsilon_p^{\rm max}$.
The optically-bright TDE remains inefficient for $pp$ interactions, whereas in the denser IR-only environment $f_{pp}$ increases and reaches $f_{pp}\simeq1$ once radiative compression enhances the target density. The maximum proton energy in the IR-only model reaches larger values at early times but decreases rapidly as the shock decelerates and radiative cooling becomes important.
}
\end{figure}

The corresponding $pp$ interaction timescale can be calculated using $t_{pp} = 1/\left( n_{\rm target}\sigma_{pp}\kappa_{pp}c \right) 
\simeq 5.6 \times 10^{8}\ {\rm s}\ \big(n_{\rm CNM}/10^{6}\ {\rm cm^{-3}}\big)^{-1}$. To account for the competition between $pp$ interactions, escape, and proton synchrotron cooling, we define the effective proton residence time as $t_{\rm eff}^{-1} = t_{\rm esc}^{-1} + t_{pp}^{-1} + t_{p,\rm syn}^{-1}$. Here we use the same effective escape time as in Equation~\eqref{eq:epmax}, treating it as a one-zone estimate of the CR residence time in the shocked region. We therefore do not distinguish between escape from the acceleration and interaction regions; in the advection-dominated limit, this prescription reduces to $t_{\rm esc} \simeq t_{\rm dyn}$. The effective $pp$ interaction efficiency can then be defined as
\be
\label{eq:fpp}
f_{pp}(\varepsilon_p,t) = \frac{t_{\rm eff}(\varepsilon_p,t)}{t_{pp}(t)}\,.
\ee
Before radiative compression becomes important and in the optically thin, advection-dominated limit, $t_{\rm eff} \simeq t_{\rm dyn}$. Therefore, we have
\be
f_{pp} \simeq \frac{t_{\rm dyn}}{t_{pp}} \simeq 0.14 \left(\frac{n_{\rm CNM}}{10^{6}\ {\rm cm^{-3}}}\right) \left(\frac{t}{1\ {\rm yr}}\right)\,,
\ee
such that $f_{pp}\propto n_{\rm CNM}t$. The resulting evolution of $f_{pp}$, evaluated at the instantaneous maximum proton energy, $\varepsilon_p^{\rm max}(t)$, is shown in Figure~\ref{fig:epmax_fpp}. In the advection-dominated limit above, $f_{pp}$ is independent of the proton energy since neither $t_{\rm dyn}$ nor $t_{pp}$ depends on it. In the full calculation, however, energy-dependent escape and proton synchrotron losses are included through $t_{\rm eff}$. For the optical TDEs, $f_{pp}$ initially decreases as the shock propagates through the steep inner CNM and subsequently increases as the shock reaches the approximately constant-density component. Nevertheless, the optical TDEs remain inefficient for $pp$ interactions over the evolution considered here. The behavior is qualitatively different for the IR-only TDEs. Because of their much larger ambient density, the $pp$ interaction time is substantially shorter and $f_{pp}$ increases with time. Once the shock becomes radiative, the enhanced target density in the compressed downstream material sharply reduces $t_{pp}$, producing the rapid increase in $f_{pp}$ seen in Figure~\ref{fig:epmax_fpp} and driving the source toward the $pp$ calorimetric regime.

The resulting maximum proton energies ($\varepsilon_p^{\rm max}$) are shown in Figure~\ref{fig:epmax_fpp}. For the IR-only TDE, $\varepsilon_p^{\rm max}$ initially increases and reaches $\sim {\rm few}\times10^8\ {\rm GeV}$ at
$t \sim 10^7\ {\rm s}$, before decreasing as the shock decelerates. The decline becomes considerably steeper at late times, when radiative cooling becomes dynamically important and the shock transitions to the momentum-conserving evolution described in Section~\ref{subsec:cool_shock}. The resulting stronger shock deceleration, together with the cooling-limited scaling $\varepsilon_{p,\rm ff}^{\rm max} \propto v_{\rm sh}^{4} n_{\rm CNM}^{-1/2}$, leads to the rapid decrease in $\varepsilon_p^{\rm max} \propto t^{-3}$. In contrast, the optical TDE model exhibits a more gradual evolution of $\varepsilon_p^{\rm max}$ and remains capable of accelerating protons to $\gtrsim 10^7\ {\rm GeV}$ over a substantial fraction of the evolution.
\subsection{Neutrino spectra and fluence}
The differential injection rate of CR protons is taken to be a power law with an exponential cutoff,
\be
Q_p(\varepsilon_p) \equiv \frac{d\dot N_p^{\rm inj}}{d\varepsilon_p} = \mathcal{N}\varepsilon_p^{-2} \exp\left( -\frac{\varepsilon_p}{\varepsilon_p^{\rm max}} \right)\,,
\ee
where the normalization $\mathcal{N}$ is determined by
\be
\int_{\varepsilon_p^{\rm min}}^\infty d\varepsilon_p\ \varepsilon_pQ_p(\varepsilon_p) = L_{\rm CR}\,,
\ee
and we set $\varepsilon_p^{\rm min}=1\ {\rm GeV}$.

The same escape and energy-loss processes determine the
steady-state proton distribution. We approximate the effective
residence time as $t_{\rm eff}^{-1} = t_{\rm esc}^{-1} + t_{pp}^{-1} + t_{p,\rm syn}^{-1}$, where $t_{pp}$ is evaluated using the target density relevant to the shocked material through which the accelerated CRs propagate. The quasi-steady-state proton distribution is then $N_p(\varepsilon_p,t) = Q_p(\varepsilon_p,t) t_{\rm eff}(\varepsilon_p,t)$. For the IR-only model in the radiative phase, this means that $t_{pp}$ in computing $t_{\rm eff}$ is evaluated using
$n_{\rm cool}$, whereas the acceleration constraint in Equation~\eqref{eq:epmax} uses $t_{pp,\rm acc}$ evaluated with the immediate post-shock density.

\begin{figure*}
\centering
\includegraphics[width=0.49\textwidth]{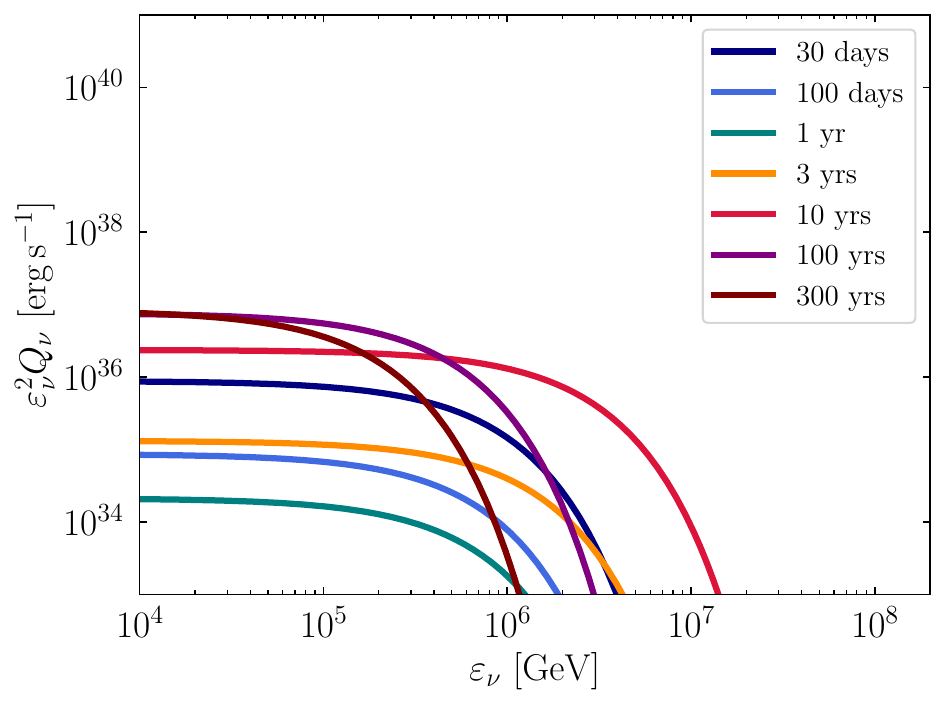}
\includegraphics[width=0.49\textwidth]{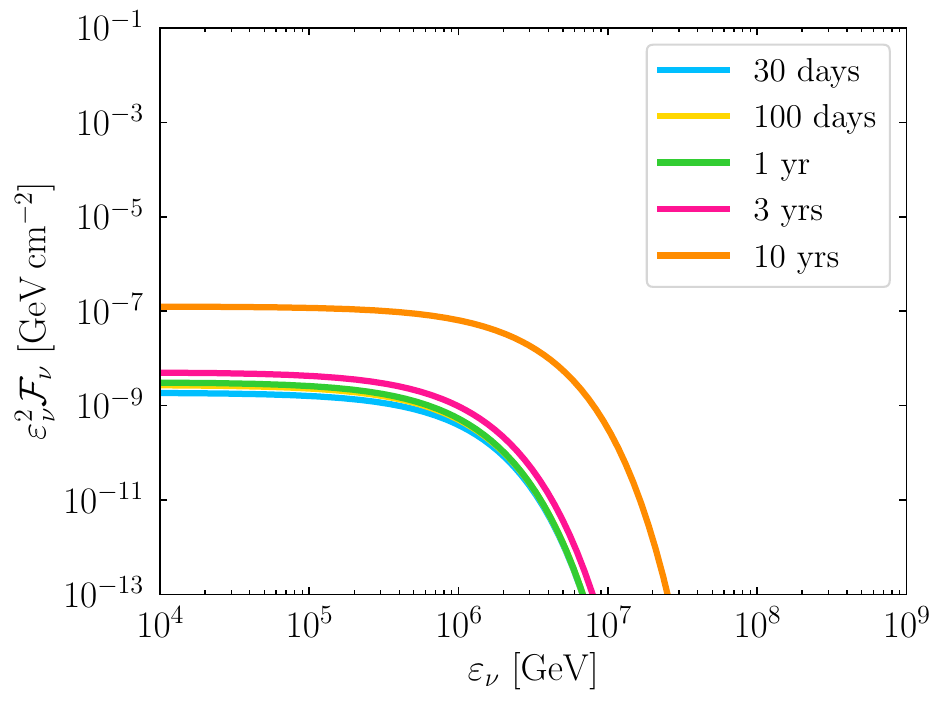}
\includegraphics[width=0.49\textwidth]{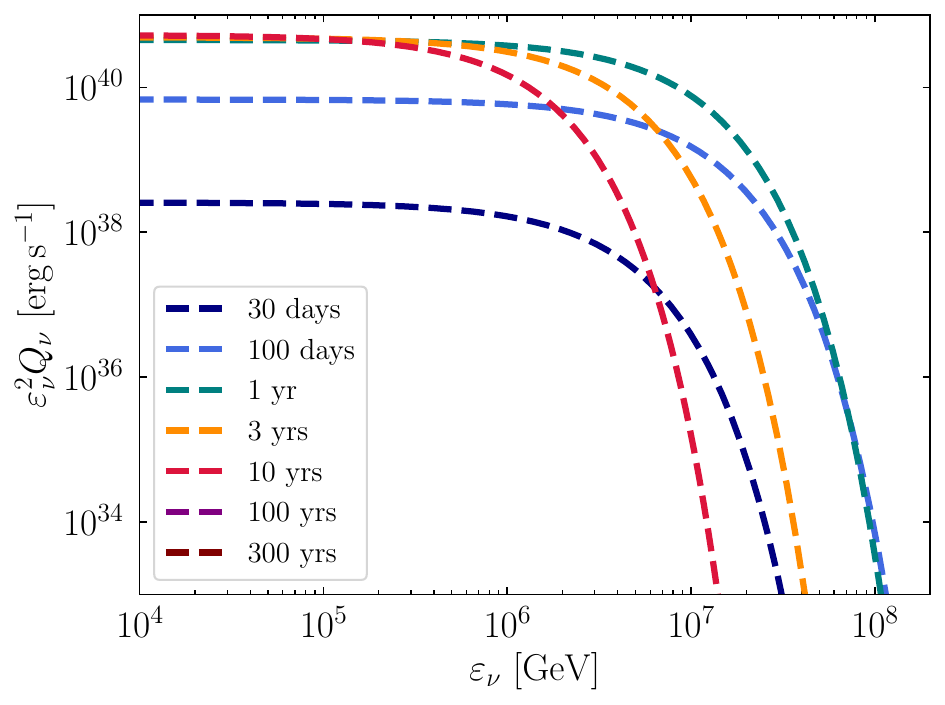}
\includegraphics[width=0.49\textwidth]{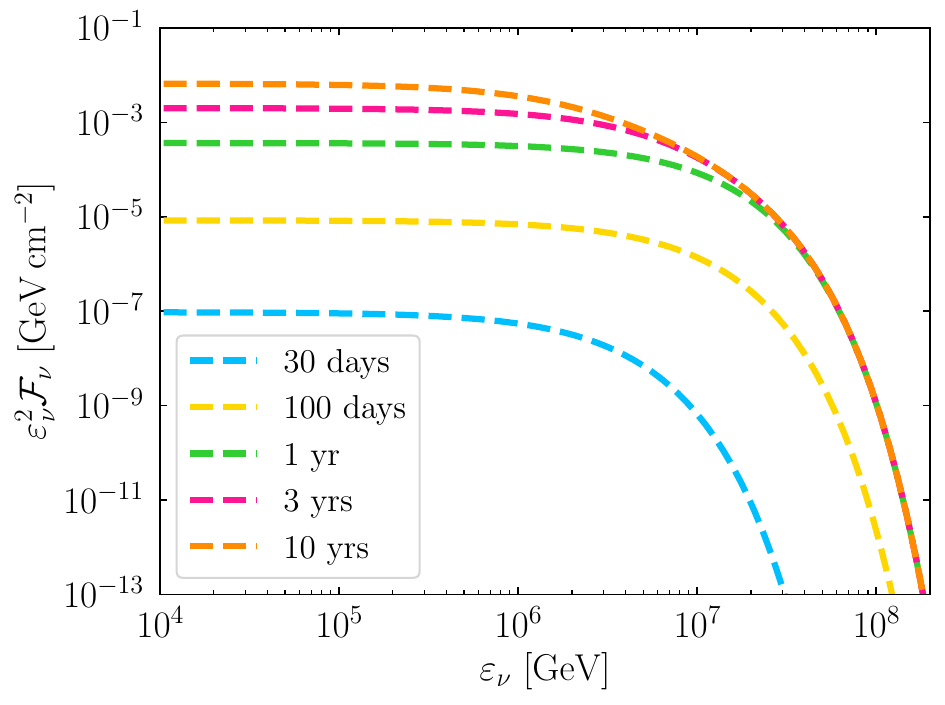}
\caption{\label{fig:nu_spectra}
Instantaneous all-flavor neutrino spectral luminosities
\emph{(left)} and time-integrated all-flavor neutrino fluences
(\emph{right}, Equation~\ref{eq:nu_flu}) for the optical \emph{(top)} and IR-only \emph{(bottom)} TDE models. The instantaneous spectra are shown
at representative times after launch of the outflow, while the
fluences are calculated for a source at $d_L=100\ {\rm Mpc}$ and accumulated over the indicated time windows ($t_{\rm obs}$). The denser IR-only environment results in substantially more efficient $pp$ neutrino production. The evolving maximum proton energy determines the high-energy cutoff of the instantaneous spectra and accumulated fluences. Note that for the IR-only TDE model, the $t = 100$ and $300$ years instantaneous spectra are absent from the bottom left panel because by these late times, the declining $\varepsilon_p^{\rm max}$ shifts the neutrino spectrum below the plotted energy range.
}
\end{figure*}

The neutrino production spectrum from $pp$ interactions is
calculated from the steady-state proton distribution. Using the
approximate relation $\varepsilon_p\simeq20\varepsilon_\nu$, we obtain
\be
\label{eq:nu_spec}
\varepsilon_\nu^2 Q_\nu(\varepsilon_\nu,t) \simeq \frac{1}{2} \left. \frac{ \varepsilon_p^2N_p(\varepsilon_p,t)}{t_{pp}(t)} \right|_{\varepsilon_p=20\varepsilon_\nu}\,,
\ee
where $Q_\nu$ denotes the all-flavor neutrino production rate.
The numerical prefactor approximately accounts for the fraction
of the parent proton energy transferred to neutrinos in $pp$ interactions. We assume standard neutrino flavor mixing, giving an approximately equal flavor ratio at Earth~\citep{Song:2020nfh}. The neutrino luminosity shown in Figure~\ref{fig:luminosity} is obtained by integrating the
all-flavor neutrino spectrum over energy,
\be
\label{eq:lnu}
L_\nu(t) = \int d (\ln \varepsilon_\nu)\ \varepsilon_\nu^2 Q_\nu(\varepsilon_\nu,t)\,.
\ee
The substantially larger $pp$ interaction efficiency in the
IR-only TDE produces a neutrino luminosity several orders of
magnitude larger than in the optical TDE case. At late times, as the IR-only model approaches the calorimetric regime, $L_\nu$ becomes increasingly controlled by the declining CR injection power, leading to the corresponding decline in the neutrino luminosity.

In the optically thin, advection-dominated limit, $t_{pp}\gg t_{\rm dyn}$ and $t_{\rm eff}\simeq t_{\rm dyn}$, so that $\varepsilon_\nu^2 Q_\nu \propto L_{\rm CR} (t_{\rm dyn}/t_{pp}) \simeq L_{\rm CR}f_{pp}$. This relation describes the qualitative behavior of the optical TDE model. During its early free-expansion phase, $L_{\rm CR} \propto t^{-1/2}$ and $f_{pp} \propto t^{-3/2}$, giving $\varepsilon_\nu^2 Q_\nu \propto t^{-2}$. After the density profile flattens, the increasing shock power and interaction efficiency cause the neutrino luminosity to rise. At late times, the energy-conserving constant-density scalings give
$L_{\rm CR} \propto t^{-1}$ and $f_{pp} \propto t$, so that the low-energy neutrino luminosity approaches a much more slowly varying evolution while the source remains optically thin.

The same optically thin scaling does not apply once the IR-only
model becomes calorimetric. When $t_{pp} \ll t_{\rm esc}$ and proton synchrotron losses are subdominant, we have $t_{\rm eff} \simeq t_{pp}$. Consequently, $N_p/t_{pp} \simeq Q_p$ and the neutrino production becomes approximately proportional to the instantaneous CR injection power rather than continuing to increase with $f_{pp}$. Thus,
once $f_{pp} \simeq 1$, increasing the target density further does not produce a corresponding increase in the bolometric neutrino conversion efficiency. Instead, the subsequent evolution is controlled primarily by the declining shock power and by the evolution of $\varepsilon_p^{\rm max}$. Therefore Figure~\ref{fig:epmax_fpp} illustrates that in the dense IR-only environment, the increasing $pp$ efficiency allows the source to approach the
calorimetric regime, while the stronger shock deceleration and eventual radiative evolution simultaneously reduce the maximum energy of the accelerated protons.

The instantaneous neutrino flux observed from a source at
luminosity distance $d_L$ is given by, $\varepsilon_\nu^2\phi_\nu \approx \varepsilon_\nu^2 Q_\nu(\varepsilon_\nu,t)/(4 \pi d_L^2)$. In the optically thin, advection-dominated limit, the instantaneous neutrino flux from a source at $d_L$ can be estimated as
\begin{align}
\varepsilon_\nu^2\phi_\nu &\simeq
\frac{\eta_{\rm CR}L_{\rm sh}f_{pp}}
{8\pi d_L^2 \ln(\varepsilon_p^{\rm max}/\varepsilon_p^{\rm min})}\\
&\simeq 2.1\times10^{-11}\ {\rm GeV\ cm^{-2}\ s^{-1}}\left(\frac{\eta_{\rm CR}}{0.1}\right)
\left(\frac{E_{\rm kin}^{\rm init}}{4.5\times10^{51}\ {\rm erg}}\right) \nonumber\\
&\times \left(\frac{n_{\rm CNM}}{10^{6}\ {\rm cm^{-3}}}\right)
\left(\frac{\ln(\varepsilon_p^{\rm max}/
\varepsilon_p^{\rm min})}{18.4}\right)^{-1}
\left(\frac{d_L}{100\ {\rm Mpc}}\right)^{-2}\,\nonumber.
\end{align}
Interestingly, the explicit time dependence cancels in this
limit, since $L_{\rm sh}\propto t^{-1}$ while
$f_{pp}\propto t$. This scaling ceases to apply once the source approaches the calorimetric regime, $f_{pp}\sim1$, in which case the neutrino luminosity instead follows the declining CR injection power.

The time-integrated neutrino fluence accumulated up to an observation time $t_{\rm obs}$ is given by
\be
\label{eq:nu_flu}
\varepsilon_\nu^2\mathcal{F}_\nu (\varepsilon_\nu;t_{\rm obs}) = \frac{1}{4\pi d_L^2} \int_{t_{\rm min}}^{t_{\rm obs}} dt\ \varepsilon_\nu^2Q_\nu(\varepsilon_\nu,t)\,,
\ee
where $t_{\rm min} = 10^6\ {\rm s}$ denotes the beginning of the
evolution, as discussed in Section~\ref{subsec:cnm}.

The resulting instantaneous fluxes and time-integrated fluences
are shown in Figure~\ref{fig:nu_spectra}. For the optical TDE, the instantaneous neutrino luminosity initially decreases as the shock propagates through the steep inner CNM and subsequently increases as the approximately constant-density component becomes
important. At late times, the declining shock power is partially compensated by the increasing $pp$ interaction efficiency. The IR-only TDE produces a substantially larger neutrino luminosity because of its much higher target density and rapidly approaches
the calorimetric regime. At later times, however, shock deceleration and radiative cooling cause the maximum proton energy to decrease, shifting the high-energy cutoff of the neutrino spectrum toward lower energies.

The time-integrated fluence reflects the cumulative effect of this evolving emission. Increasing the integration window therefore increases the total fluence, but the gain depends strongly on the epoch over which the signal is accumulated. In particular, late-time emission can continue to increase the low-energy fluence even after the maximum neutrino energy has decreased substantially. At the same time, longer integration windows accumulate more background events in a neutrino search. The optimal search window is therefore determined by the competition between the cumulative signal and the corresponding background, motivating searches over multiple physically motivated time windows, which we shall discuss in Section~\ref{sec:search}.
\subsection{Contribution to the all-sky diffuse neutrino flux}
\label{subsec:diffuse}
\begin{figure}
\centering
\includegraphics[width=0.48\textwidth]{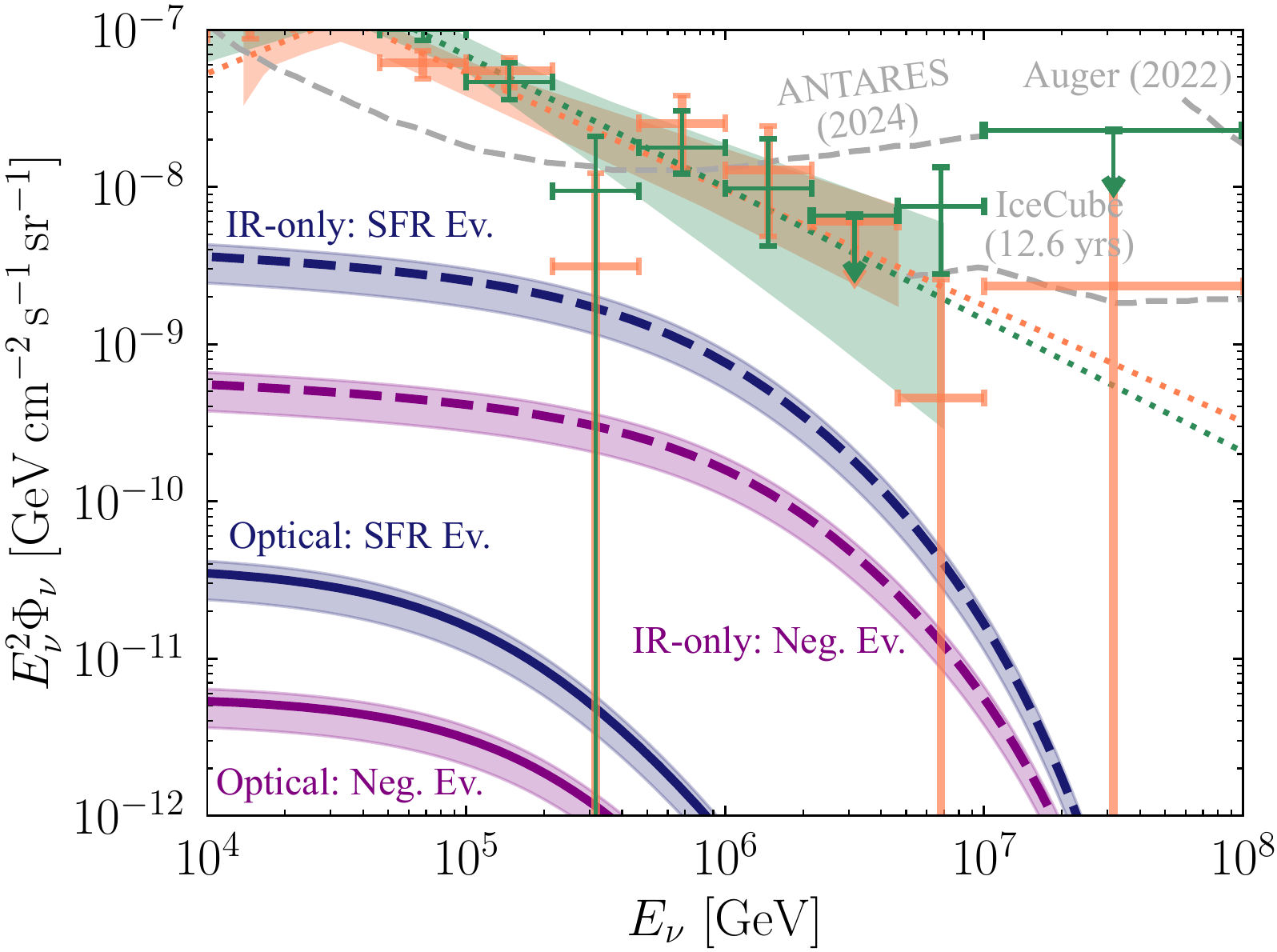}
\caption{\label{fig:diffuse}Contribution of high-energy neutrinos from TDE-driven outflow-CNM interactions to the all-sky diffuse neutrino intensity. The predictions are shown separately for the optically-bright and IR-only TDE populations, for the negative TDE-evolution and SFR-tracing scenarios considered in this work. For the optically-bright population, we adopt the local volumetric
rate $\dot{\rho}_{\rm opt,0} = 3.1^{+0.6}_{-1.0}\times10^{-7}\ {\rm Mpc^{-3}\,yr^{-1}}$ inferred by~\cite{Yao:2023rbr}. For the IR-only population, we adopt $\dot{\rho}_{\rm IR,0}=\dot{\rho}_{\rm opt,0}/3$, motivated by the somewhat lower local rate inferred for IR-only TDEs by~\cite{Masterson:2024rur}. For each event, the neutrino emission is integrated from $t_{\rm min}=10^6\ {\rm s}$ to $t_{\rm end} = 10^{10}\ {\rm s}\simeq3.2\times10^2\ {\rm yr}$. The diffuse upper limits from ANTARES at 95\% C.L.~\citep{ANTARES:2024ihw}, Auger at 90\% C.L.~\citep{PierreAuger:2023pjg}, and IC-EHE at 90\% C.L.~\citep{IceCube:2018fhm} are shown as dashed gray lines. The orange and green shaded regions show the $68\%$ C.L. uncertainties of the IceCube broken power-law (BPL) fits from the Combined Fit (CF) and Medium Energy Starting Events (MESE) analyses, respectively, with the corresponding best-fit spectra shown as dotted lines~\citep{IceCube:2025tgp}. The segmented flux measurements from the CF and MESE analyses are shown by the orange and green crosses, respectively. The reported per-flavor IceCube fluxes have been multiplied by a factor of three, assuming a $1:1:1$ flavor ratio at Earth, for comparison with the all-flavor TDE results shown here.
}
\end{figure}
In this subsection, we estimate the contribution of TDE-driven outflow-CNM interactions to the all-sky diffuse high-energy neutrino intensity. We calculate the contributions from the optical and IR-only TDE populations separately, using the corresponding single-source neutrino yields derived in the previous subsection. This distinction is particularly important because the much denser environment of the IR-only population leads to substantially more efficient $pp$ interactions and hence a larger neutrino yield per event, while the inferred local occurrence rate of IR-selected TDEs is somewhat lower than that of optically selected TDEs.

In contrast to the finite observational time windows considered
in Figure~\ref{fig:nu_spectra}, the diffuse calculation requires
the time-integrated neutrino yield of an individual event. Since
our primary interest is in the high-energy neutrino emission, we
follow the shock evolution up to $t_{\rm end} = 10^{10}\ {\rm s}\simeq 3.2 \times 10^2\ {\rm yr}$, by which time the contribution to the high-energy neutrino yield has become negligible. The time-integrated source-frame neutrino spectrum entering the diffuse calculation is therefore
\be
\label{eq:source_yield}
\frac{dN_{\nu,j}}{dE_\nu'} = \int_{t_{\rm min}}^{t_{\rm end}} dt\ Q_{\nu,j}(E_\nu',t)\,,
\ee
where $j=\{{\rm opt},{\rm IR}\}$ denotes the optical and IR-only TDE populations, respectively.

For a TDE population with comoving volumetric event-rate density
$\dot{\rho}_j(z)$, the corresponding all-flavor diffuse neutrino
intensity is
\be
\label{eq:diffuse_flux}
E_\nu^2\Phi_{\nu,j}(E_\nu) = \frac{c}{4\pi} \int_0^{z_{\rm max}} \frac{dz}{H(z)}\ \dot{\rho}_j(z)\ E_\nu^2 \left. \frac{dN_{\nu,j}}{dE_\nu'} \right|_{E_\nu'=(1+z)E_\nu}\,.
\ee
Here $\dot{\rho}_j(z)$ is defined per unit comoving volume and
source-frame time, $E_\nu'=(1+z)E_\nu$ is the source-frame neutrino energy, and we adopt $z_{\rm max} = 6$, beyond which the contribution is negligible
for the source-evolution models considered here. For a flat $\Lambda$CDM cosmology, $H(z) = H_0 \sqrt{\Omega_m(1+z)^3+\Omega_\Lambda}$. We parameterize the redshift-dependent rate of each population as
\be
\label{eq:tde_rate}
\dot{\rho}_j(z) = \dot{\rho}_{j,0}f(z)\,,
\ee
where $\dot{\rho}_{j,0}$ is its local volumetric rate and
$f(z)$ describes the cosmological evolution, normalized such that
$f(0)=1$. We adopt the general smoothly joined power-law form
\be
\label{eq:fz}
f(z) = \mathcal{A} \left[ (1+z)^{a\eta} + \left(\frac{1+z}{B}\right)^{b\eta}
+ \left(\frac{1+z}{C}\right)^{c\eta} \right]^{1/\eta}\,,
\ee
where $\mathcal{A}$ is chosen such that $f(0)=1$. The parameters adopted for the two redshift-evolution scenarios considered in this work are summarized in Table~\ref{tab:tde_evolution}.

\begin{table}
\centering
\caption{Parameters describing the redshift evolution of the TDE
rate density considered in this work.}
\label{tab:tde_evolution}
\begin{tabular}{lcccccc}
\hline
\hline
Evolution & $a$ & $b$ & $c$ & $B$ & $C$ & $\eta$ \\
\hline
Negative & $0.2$ & $-3.2$ & $-7.0$ & $1.43$ & $2.66$ & $-2$ \\
SFR      & $3.4$ & $-0.3$ & $-3.5$ & $5\times10^3$ & $9$ & $-10$ \\
\hline
\end{tabular}
\end{table}
The cosmological evolution of the TDE population remains uncertain, and we therefore consider the following two representative prescriptions:

\begin{itemize}
\item \emph{Negative evolution:} We adopt the TDE redshift evolution derived by~\cite{Sun:2015bda}. Assuming that the average TDE rate per galaxy does not evolve with redshift, they relate the volumetric TDE rate to the evolving number density of SMBHs. Their empirical fit is
described by Equation~\eqref{eq:fz} with the parameters
listed in Table~\ref{tab:tde_evolution}. This prescription gives a TDE
rate density that decreases with increasing redshift.
\item \emph{SFR evolution:} As an alternative scenario, we assume that the TDE rate traces the redshift evolution of the cosmic star-formation rate (SFR). We adopt the smoothly broken power-law parameterization of~\cite{Yuksel:2008cu}, corresponding to the parameters listed in Table~\ref{tab:tde_evolution}. In this case the TDE rate increases rapidly with redshift at low $z$, approximately as $(1 + z)^{3.4}$, evolves only weakly between $z \simeq 1$ and $4$, and decreases at higher $z$. We use only the redshift dependence of the SFR to describe the TDE population; its overall normalization is independently fixed by the local TDE rate $\dot{\rho}_{\rm TDE,0}$.
\end{itemize}
For the optical TDE population, we adopt the local volumetric rate inferred by~\cite{Yao:2023rbr}, $\dot{\rho}_{\rm opt,0} = 3.1^{+0.6}_{-1.0} \times 10^{-7}\ {\rm Mpc^{-3}\,yr^{-1}}$, for optically selected TDEs with
$L_{\rm bb} > 10^{43}\ {\rm erg\ s^{-1}}$, where $L_{\rm bb}$ is the bolometric blackbody luminosity of the optical/UV TDE emission. We emphasize that the optical luminosity criterion used to define this TDE population does not determine the kinetic energy of the outflow in our model, which is instead independently specified by the adopted $M_{\rm ej}$ and $v_{\rm sh,init}$. We adopt the same outflow parameters, and hence the same initial kinetic energy, for the optical and IR-only TDE populations.  For the IR-only population,~\cite{Masterson:2024rur} inferred a local
IR-selected TDE rate of $\simeq 1.3 \times 10^{-7}\ {\rm Mpc^{-3}\ yr^{-1}}$, which is smaller than the optical TDE rate, while emphasizing that
the inferred IR rate should be regarded as a lower limit because
of incompleteness in the IR-selected sample. Motivated by this
comparison, we adopt the simple benchmark $\dot{\rho}_{\rm IR,0} = (1/3) \dot{\rho}_{\rm opt,0}$, for the IR-only population.

The resulting diffuse neutrino intensities are shown in
Figure~\ref{fig:diffuse}. Two effects determine the relative
contributions of the optical and IR-only populations - their local occurrence rates and their neutrino yields per event. Although we adopt a smaller local rate for the IR-only population, its much denser CNM environment produces substantially more efficient $pp$ interactions. As shown in Figure~\ref{fig:epmax_fpp}, the IR-only model can approach the
$pp$ calorimetric regime, whereas the optical TDE model remains optically thin with respect to $pp$ interactions. Consequently, the larger single-event neutrino yield of the IR-only population can more than compensate for its lower assumed occurrence rate.

For either population, the SFR-tracing evolution produces a larger diffuse intensity than the negative-evolution prescription. This difference follows from the cosmological weighting in Equation~\eqref{eq:diffuse_flux}. In the negative-evolution scenario, the TDE rate decreases with increasing redshift, so the diffuse intensity is weighted predominantly toward the nearby Universe. In the SFR-tracing scenario, the source density instead increases strongly toward $z\sim1$ and remains comparatively large to higher redshifts, increasing the integrated cosmological contribution.

The difference between the two evolution scenarios is also energy dependent. Neutrinos observed at $E_\nu$ are emitted at $E_\nu'=(1+z)E_\nu$, so at sufficiently high observed energies the contribution from high-redshift sources probes progressively higher energies in the single-event spectrum. As the source-frame energy approaches the spectral cutoff, the high-redshift contribution is increasingly suppressed. The precise spectral shape of the diffuse intensity therefore reflects both the cosmological source evolution and the time-integrated neutrino spectrum of each TDE population.

At $E_\nu=100\ {\rm TeV}$, the predicted all-flavor diffuse neutrino intensities from the optical TDE population are $E_\nu^2 \Phi_{\nu,\rm opt} \simeq 3.4\times10^{-12}$ and $1.7\times10^{-11}\ {\rm GeV\ cm^{-2}\ s^{-1}\ sr^{-1}}$ for the negative- and SFR-evolution scenarios, respectively. The corresponding intensities from the IR-only population are
substantially larger, $E_\nu^2 \Phi_{\nu,\rm IR} \simeq 4.3 \times 10^{-10}$ and $2.6 \times 10^{-9}\ {\rm GeV\ cm^{-2}\ s^{-1}\ sr^{-1}}$.
Thus, despite adopting a local IR-only TDE rate that is a factor of three smaller than that of the optical TDE population, the diffuse neutrino intensity is dominated by the IR-only TDEs because of their substantially larger single-source neutrino yields.

For comparison, the recent IceCube Combined Fit (CF) measurement yields a per-flavor diffuse astrophysical neutrino intensity of $E_\nu^2\Phi_{\nu,\rm astro} \simeq 1.77 \times 10^{-8}\ {\rm GeV cm^{-2} s^{-1} sr^{-1}}$ at $E_\nu = 100\ {\rm TeV}$~\citep{IceCube:2025tgp}. Assuming an approximately $1:1:1$ flavor ratio at Earth, this corresponds to an all-flavor intensity of $ \simeq 5.3 \times 10^{-8}\ {\rm GeV cm^{-2} s^{-1} sr^{-1}}$. For the central local TDE rates adopted here, the combined optical and IR-only populations yield $E_\nu^2\Phi_{\nu,\rm tot} \simeq 4.3 \times 10^{-10}$ and $2.6 \times 10^{-9}\ {\rm GeV\ cm^{-2}\ s^{-1}\ sr^{-1}}$, corresponding to approximately $1\%$ and $5\%$ of the measured all-flavor diffuse intensity for the negative- and SFR-evolution scenarios, respectively. This contribution is overwhelmingly dominated by the IR-only population, where the optical TDEs alone account for only $\sim 0.007\%$ - $0.03\%$, whereas IR-only TDEs contribute approximately $0.9\%$ - $5\%$ for the central rate normalizations. Accounting for the uncertainty in the local TDE rate, the IR-only contribution can approach the $\sim 10\%$ level in the more strongly evolving population scenario. Thus, although optically-bright TDE outflow-CNM interactions make only a negligible contribution to the diffuse neutrino background, TDEs embedded in dense, obscured environments can constitute a
significant component of the observed high-energy neutrino intensity, despite their lower local occurrence rate.
\section{Search strategy and prospects in current and future neutrino telescopes}
\label{sec:search}
Having discussed the plausible high-energy neutrino signatures from shocked CNM around TDEs and their contribution to the diffuse flux, in this section we explore the detection prospects at IceCube, IceCube-Gen2, KM3NeT, and HUNT. We focus here on the IR-only TDE population, for which the dense circumnuclear environments yield the most favorable prospects for high-energy neutrino searches.
\subsection{Stacking strategy and TDE population}
\label{subsec:stacking}
The expected number of neutrino events from an individual TDE is small, motivating a stacking search using a catalog of electromagnetically identified TDEs. Such a search can exploit both the known source positions and the temporal association with the TDE, thereby substantially reducing the neutrino background relative to an untriggered all-sky search.

We follow the general stacking formalism developed in Appendix D of~\cite{Mukhopadhyay:2026afk}, to which we refer the reader for details, and summarize the relevant ingredients here. We consider a population of TDEs extending to a maximum redshift $z_{\rm max}$, and associate with each source a neutrino search window of duration $\delta t_{\rm search}$ following the TDE. The expected signal is obtained by folding the redshifted time-integrated neutrino fluence with the detector effective area and summing over the TDE population. The corresponding background is accumulated over the same temporal and angular search windows, and therefore increases with both the number of stacked sources and $\delta t_{\rm search}$.

The number of TDEs entering the stacking analysis follows directly from the cosmological TDE rate introduced in Section~\ref{subsec:diffuse}. For an observation time $T_{\rm obs}$, the expected number of TDEs within a maximum redshift $z_{\rm max}$ is
\be
\label{eq:ntde}
N_{\rm TDE}(z_{\rm max}, T_{\rm obs}) = T_{\rm obs} \int_0^{z_{\rm max}} dz\ \frac{\dot{\rho}_j(z)}{1+z} \frac{dV_{\rm com}}{dz},
\ee
where $\dot{\rho}_j(z)$ is given by Equation~\eqref{eq:tde_rate}, $dV_{\rm com}/dz$ is the differential comoving volume, and the factor $(1+z)^{-1}$ accounts for cosmological time dilation.

\begin{figure}
\centering
\includegraphics[width=0.48\textwidth]{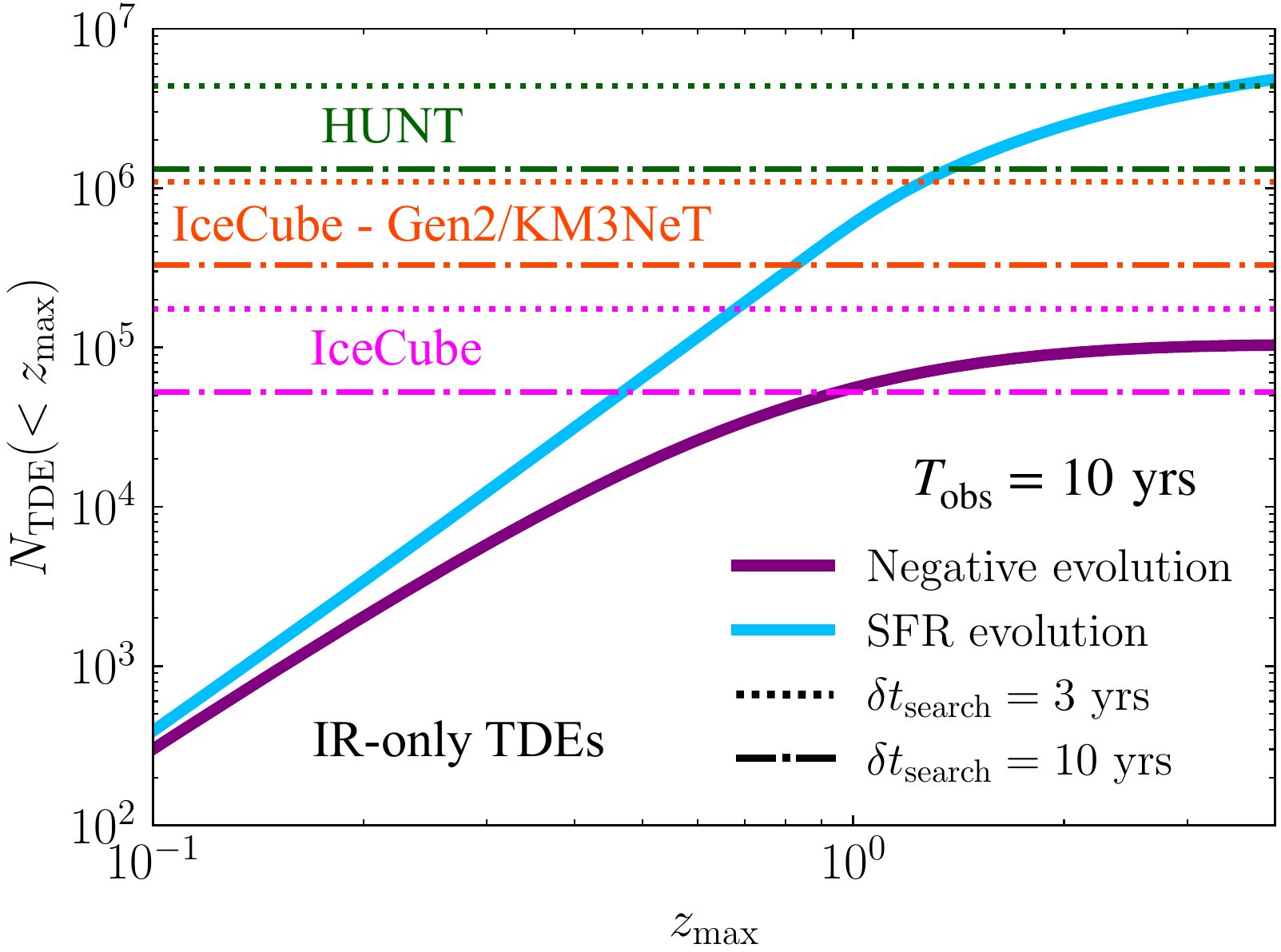}
\caption{\label{fig:ntde_zmax}Cumulative number of IR-only TDEs within a maximum redshift $z_{\rm max}$ over an observation time $T_{\rm obs} = 10\ {\rm yrs}$. The purple and blue curves correspond to the negative and SFR-tracing TDE evolutions, respectively. The horizontal lines indicate the characteristic source overlap scale ($N_{\rm TDE}^{\rm overlap}$ see Equation~\ref{eq:ntde_overlap}) for IceCube (magenta), IceCube-Gen2 (orange red), KM3NeT (orange red), and HUNT (dark green), assuming search windows of $\delta t_{\rm search}=3\ {\rm yrs}$ (dotted) and $10\ {\rm yrs}$ (dash-dotted).
}
\end{figure}

In Figure~\ref{fig:ntde_zmax} we show the cumulative number of IR-only TDEs expected within $z_{\rm max}$ over $T_{\rm obs}=10\ {\rm yrs}$. At low redshift, the predictions for the two redshift evolutions are similar, whereas they increasingly diverge toward higher redshifts. The SFR-tracing evolution predicts a substantially larger number of high-redshift TDEs, while the negative evolution leads to a comparatively slower increase in the cumulative source population.

For the neutrino search, we consider IceCube, IceCube-Gen2, KM3NeT, and
HUNT. For each detector, we adopt the corresponding muon-neutrino effective area $A_{\rm eff}(E_\nu)$ and a characteristic angular uncertainty $\theta_{\rm err}$. We restrict our analysis to track-like events induced by muon neutrinos and antineutrinos, which provide the best angular resolution for source association. For IceCube, we use the declination-dependent muon-neutrino effective area from the 14 years IceTracks-DR2 point-source data release~\citep{Abbasi:2026ehs}. We evaluate the effective area separately in the Northern and Southern hemispheres by averaging over the corresponding declination ranges (see, e.g.,~\citealt{Mukhopadhyay:2023niv,Matsui:2023ohr,Mukhopadhyay:2024lwq}). The IceCube-Gen2 effective areas are obtained by scaling the IceCube effective areas by a factor of $10^{2/3}$. For KM3NeT we consider the muon neutrino effective area from~\cite{KM3Net:2016zxf}. We also consider a future typical $30\ {\rm km}^3$ telescope in the form of HUNT~\citep{Chen:2026jdx} (similar proposals exist for TRIDENT~\citep{TRIDENT:2022hql} and NEON~\citep{Zhang:2024slv}). Given that HUNT will be water-based and have similar characteristics to KM3NeT, we use the KM3NeT effective area scaled by the detector volume ($\sim 30^{2/3}$) as an approximation for HUNT's effective area.

We adopt $\theta_{\rm err}=0.5^\circ$, $0.2^\circ$, $0.2^\circ$, and $0.1^\circ$ for IceCube, IceCube-Gen2, KM3NeT, and HUNT, respectively. These are typical values for ice- and water-based detectors corresponding to the angular error for tracks for current and upcoming neutrino telescopes. The corresponding angular search region is $\Delta\Omega_{\rm err} = 2 \pi \big( 1- \cos \theta_{\rm err} \big)$. As the number of stacked TDEs increases, the angular and temporal search regions associated with different sources increasingly overlap. For a population containing $N_{\rm TDE}$ events over $T_{\rm obs}$, the mean number of sources with active search windows at any given time is approximately $N_{\rm active} = N_{\rm TDE} \big(\delta t_{\rm search}/T_{\rm obs} \big)$. We define a characteristic source-overlap scale by requiring the summed angular search regions of the active sources to become comparable to the full sky, $N_{\rm active}\Delta\Omega_{\rm err}\simeq4\pi$, corresponding to
\be
\label{eq:ntde_overlap}
N_{\rm TDE}^{\rm overlap} (\delta t_{\rm search}, \Delta\Omega_{\rm err}) = \frac{T_{\rm obs}}{\delta t_{\rm search}} \frac{4\pi}{\Delta\Omega_{\rm err}}.
\ee
The horizontal lines in Figure~\ref{fig:ntde_zmax} show this scale for each detector for $\delta t_{\rm search} = 3$ and $10\ {\rm yrs}$. For the negative evolution, the cumulative number of TDEs remains below the source-overlap scale over most of the redshift range considered. In contrast, for the SFR-tracing evolution the larger high-redshift population can lead to substantial source overlap, particularly for longer search windows and detectors with poorer angular resolution. We emphasize that Equation~\eqref{eq:ntde_overlap} does not represent a sharp limit for stacking, but indicates when the search regions of different TDEs begin to substantially overlap, reducing the advantage of spatial and temporal source localization.
\subsection{Detection prospects}
\begin{figure}
\centering
\includegraphics[width=0.48\textwidth]{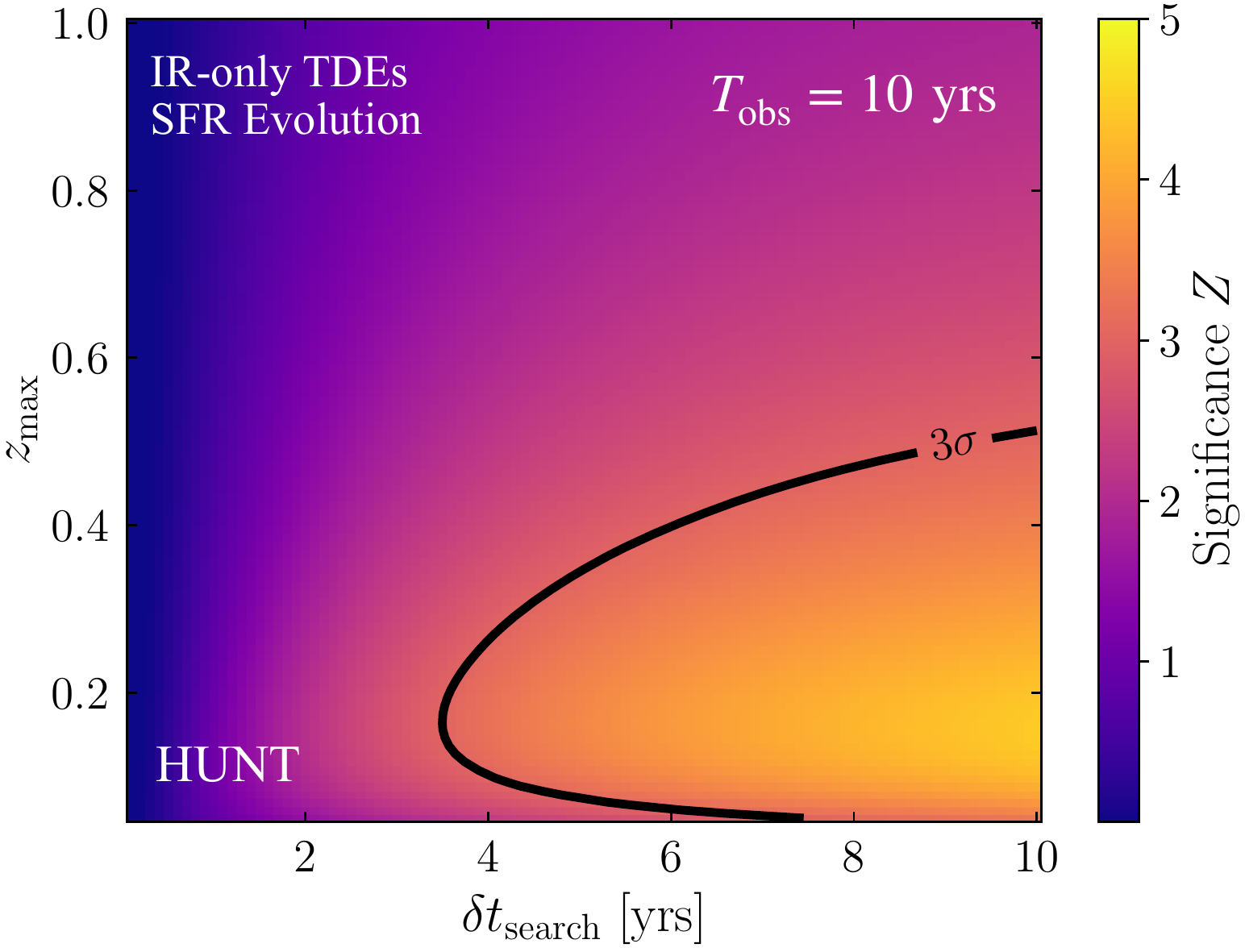}\\
\includegraphics[width=0.48\textwidth]{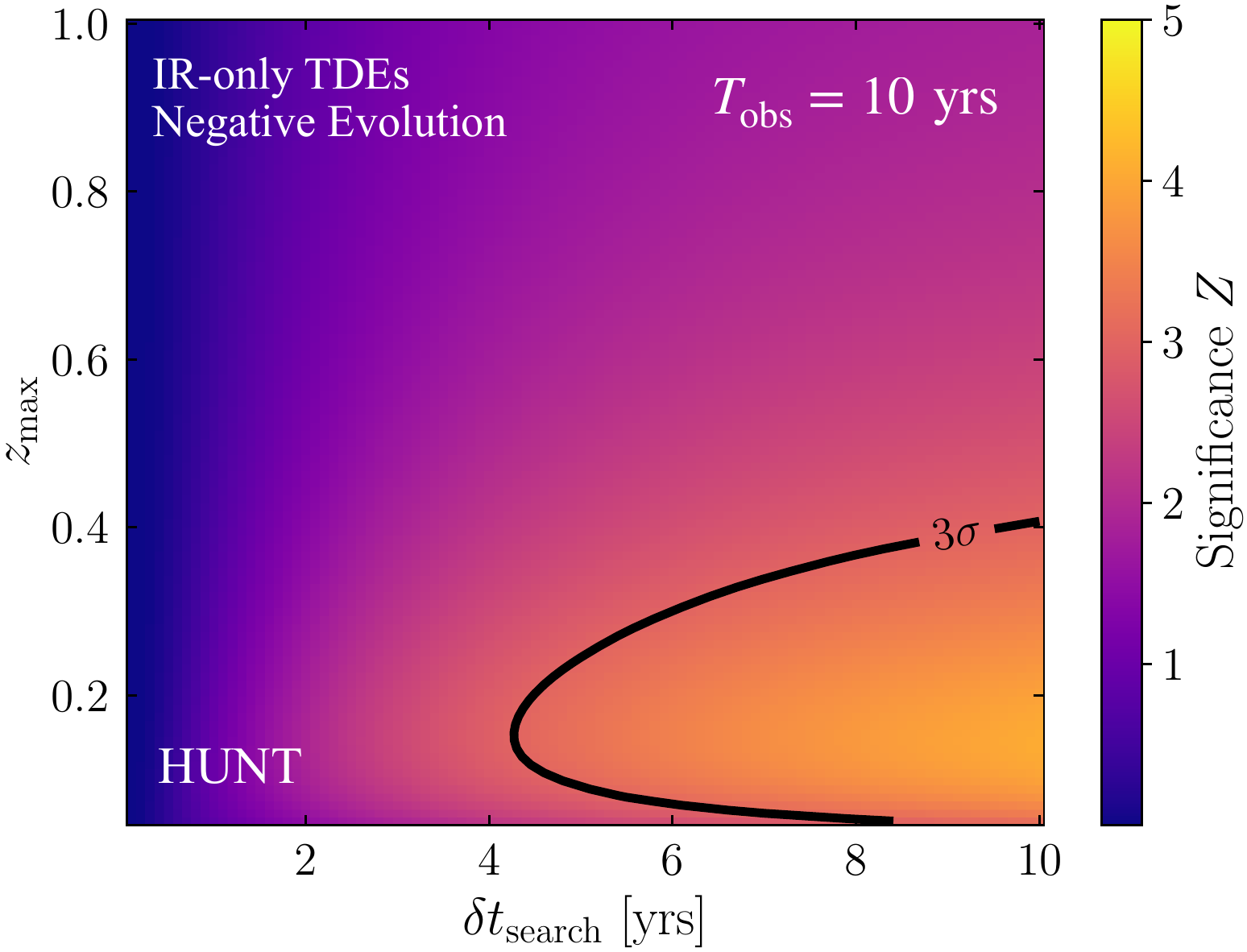}
\caption{\label{fig:hunt_sig} Expected stacked detection significance for HUNT as a function of the maximum redshift $z_{\rm max}$, and neutrino search-window duration $\delta t_{\rm search}$, for the IR-only TDE population with the fiducial value of $n_{\rm env}=10^6\ {\rm cm^{-3}}$. The \emph{top} and \emph{bottom} panels correspond to the SFR-tracing and negative TDE evolutions respectively. We adopt an observation time $T_{\rm obs}=10\ {\rm yrs}$ and consider neutrinos
with $E_\nu > 10^5\ {\rm GeV}$. The black curve denotes a
$3\sigma$ detection significance.
}
\end{figure}
Increasing either $z_{\rm max}$ or $\delta t_{\rm search}$ does not
necessarily improve the sensitivity. Extending the catalog to larger
$z_{\rm max}$ increases the number of stacked TDEs, but the contribution from individual sources decreases with distance, while the accumulated background continues to increase. Similarly, as shown in Figure~\ref{fig:nu_spectra}, the high-energy neutrino emission in the outflow-CNM scenario can persist from months to several years. A short search window therefore reduces the background but captures only part of the neutrino fluence, whereas a longer window collects more signal at the expense of additional background. We therefore explore the detection prospects in the two-dimensional $(z_{\rm max},\delta t_{\rm search})$ plane.

We calculate the stacked signal and background following Appendix D
of~\cite{Mukhopadhyay:2026afk}, with a few modifications relevant for the present analysis. We restrict the search to neutrino energies $\varepsilon_\nu>10^5\ {\rm GeV}$, where the atmospheric neutrino background is strongly suppressed. Rather than fixing $z_{\rm max}$ and $\delta t_{\rm search}$, we vary both quantities, with the neutrino fluence from each TDE integrated over the corresponding observer-frame search window. Cosmological time dilation is accounted for by evaluating the source-frame emission up to $t_{\rm src}=\delta t_{\rm search}/(1+z)$.

For our IR-only TDE benchmark with $n_{\rm env}=10^6\ {\rm cm^{-3}}$, IceCube, IceCube-Gen2, and KM3NeT remain below a $3\sigma$ detection significance over the parameter space considered. HUNT provides the most promising prospects, and in Figure~\ref{fig:hunt_sig} we show the resulting significance for $T_{\rm obs}=10\ {\rm yrs}$ assuming both the SFR-tracing (top panel) and negative (bottom panel) TDE evolutions. For both evolutions, the significance exhibits a clear optimum with $z_{\rm max}$. At small $z_{\rm max}$, increasing the catalog depth improves the sensitivity by adding nearby TDEs to the stacked signal. At larger $z_{\rm max}$, however, the additional sources are increasingly distant, and their contribution to the stacked signal no longer compensates for the accompanying increase in background. This is particularly relevant for IR-only TDEs, for which current IR-selected samples extend only to relatively low redshifts, $z \lesssim 0.4$~\citep{Necker:2024vfs,Nair:2026lfv}. Importantly, our results indicate that extending the catalog to $z \sim 1$ is not required for an optimal stacking search, which instead favors relatively nearby sources with $z_{\max} \sim 0.1 – 0.4$.

The significance also generally increases toward longer search windows over the range considered, as a larger fraction of the neutrino emission at later times is collected. The SFR-tracing evolution yields a larger significance at higher redshifts, owing to its larger TDE population, and consequently extends the $3 \sigma$ region to somewhat larger $z_{\rm max}$ than the negative evolution. Nevertheless, in both cases a $3\sigma$ detection is possible over a region of parameter space, favoring relatively nearby TDE catalogs and multi-year neutrino search windows.
\subsection{Electromagnetic surveys for TDE stacking searches}
\label{subsec:em_surveys}
The prospects for such EM-informed stacking searches will improve substantially with the rapidly growing samples of TDEs expected from wide-field time-domain surveys. In the optical, the Vera C. Rubin Observatory Legacy Survey of Space and Time (LSST) will survey approximately $18000\ {\rm deg^2}$ of the southern sky in six optical bands, with forecast studies predicting thousands of TDE detections
per year and a detected population extending to $z \sim 1$, although
the majority of events are expected at substantially smaller redshifts~\citep{2019MNRAS.488.4042T}. Complementarily, the Wide Field
Survey Telescope (WFST) will survey approximately $8000\ {\rm deg^2}$ of the northern sky, with mock-survey calculations predicting $\sim 400$ TDE detections per year for a TDE-optimized survey and detections extending to $z \sim 0.8$~\citep{Lin:2022npv}. These complementary optical coverages are particularly useful for neutrino searches. The northern sky WFST sample overlaps favorably with the region in which IceCube and IceCube-Gen2 can exploit upgoing muon tracks, whereas the predominantly southern LSST coverage is complementary to the strong southern sky sensitivity of KM3NeT/ARCA.

The recently launched Nancy Grace Roman Space Telescope will also provide an additional near-infrared view of the TDE population through its High Latitude Time Domain Survey (HLTDS). Recent forecasts suggest that the HLTDS could detect up to $\sim 100$ TDEs per year, including events at substantially higher redshifts than those accessible to wide-field optical surveys~\citep{Karmen:2026ixp}. Roman's Wide Field Instrument covers $0.48$-$2.3\ \mu{\rm m}$, such that rest-frame optical/UV emission from distant TDEs is redshifted into its near-infrared bands. This provides sensitivity to TDEs at $z>1$, with detections potentially extending to $z \gtrsim 3$~\citep{Karmen:2026ixp}. On the other hand, Roman's wavelength coverage extends only to $2.3~\mu{\rm m}$ and therefore samples primarily the short-wavelength side of the thermal dust emission associated with the infrared TDE echoes relevant to the obscured population considered here. In this respect, Roman is complementary to longer-wavelength surveys such as WISE/NEOWISE and SPHEREx, rather than directly probing the same mid-infrared-selected population. Its comparatively small time-domain survey area also makes its role distinct from the much wider optical samples expected from Rubin/LSST and WFST.

Infrared time-domain surveys provide a complementary avenue for constructing catalogs of the obscured TDE population considered here. The all-sky coverage and long temporal baseline of WISE/NEOWISE have already enabled systematic searches for obscured IR transients and subsequently IR-selected TDE candidates~\citep{Masterson:2024rur}. Looking forward, SPHEREx  is carrying out repeated all-sky spectrometric and photometric surveys over $0.75 - 5\ \mu{\rm m}$~\citep{2020SPIE11443E..0IC}, extending to wavelengths that probe the thermal dust emission relevant to IR TDE dust echoes. Such all-sky IR samples can be cross-correlated with both IceCube/IceCube-Gen2 and KM3NeT, providing a complementary route to population-resolved neutrino stacking searches for obscured IR-only TDEs. In particular, increasing the completeness of IR-selected TDE samples over the relatively nearby redshift range favored by Figure~\ref{fig:hunt_sig} may be more important for neutrino stacking searches than extending the catalogs to substantially higher redshifts.
\section{Discussion}
\label{sec:disc}
In this section, we provide some additional discussion relevant to the current work. In particular, we outline the estimate regarding the $p\gamma$ neutrino production channel, discuss the caveats of our current work, and compare it with the existing literature.
\subsection{Photohadronic ($p\gamma$) Interactions}
\label{subsec:pgamma}
In addition to the $pp$ interactions discussed in this work, the CR protons accelerated at the forward shock can also interact with the IR photons which are emitted as a result of the optical/UV photons being reprocessed by the dust around TDEs. In such a scenario the photohadronic ($p\gamma$) interactions can contribute to the production of high-energy neutrinos. In this subsection, we provide some estimates for the conditions required for the $p\gamma$ interaction channel and show that they are difficult to be realized for the sub-relativistic shocks considered here.
\subsubsection{Dust-reprocessed IR photons.} 
For an approximately thermal IR photon field with temperature $T_{\rm IR}$, the characteristic photon energy is given by $\varepsilon_\gamma^{\rm IR} \sim 2.8\ k_B T_{\rm IR} \simeq 0.24\ {\rm eV}$, where we assume typical IR photon temperature $T_{\rm IR} \sim 10^3\ {\rm K}$~\citep{Lu:2015jfg}. Therefore, the threshold CR proton energy required for $\Delta-$resonance is given by $\varepsilon_p^{\Delta-\rm res} \simeq \big(0.3\ {\rm GeV}/ \varepsilon_\gamma^{\rm IR}\big) m_p c^2 \sim 10^9\ {\rm GeV}$. Achieving such high energies for the CR protons in this scenario is difficult even assuming maximal acceleration efficiency. As shown in Figure~\ref{fig:epmax_fpp}, $\varepsilon_p^{\rm max}$ even for the dense IR-only TDE scenario barely reaches $\sim 2 \times 10^8\ {\rm GeV}$ at its peak and subsequently decreases as the shock decelerates and becomes radiative. The maximum proton energy attained in our model therefore remains below the characteristic energies required for efficient photomeson production through the $\Delta$-resonance channel. Therefore, we conclude that the $p\gamma$ interactions with the characteristic IR photons are suppressed for the sub-relativistic shocks considered here. An additional distinction between the $pp$ scenario considered here and IR-target $p\gamma$ models is the expected neutrino timing. The IR echo is delayed relative to the intrinsic optical/UV flare by the light-crossing time, $t_{\rm IR}\sim R_{\rm dust}/c$, so neutrino production in IR-target $p\gamma$ models can approximately track the delayed IR emission~\citep{Yuan:2024foi}. In our scenario, the neutrino emission instead follows the evolving outflow–CNM interaction and is not directly tied to the IR light curve.
\subsubsection{Primary optical/UV photons} 
We also note that primary optical/UV radiation from the TDE can in principle also provide targets for $p\gamma$ interactions. For characteristic photon energies $\varepsilon_\gamma^{\rm opt/UV} \sim \mathcal{O}(1 - 10\ {\rm eV})$, we have $\varepsilon_p^{\Delta-\rm res} \simeq 3 \times 10^7\ {\rm GeV}\ \big( \varepsilon_\gamma^{\rm opt/UV}/10\ {\rm eV} \big)^{-1}$. This is barely achievable for the optical TDEs (see Figure~\ref{fig:epmax_fpp}). Additionally, the CR acceleration occurs at the forward shock which is far from the central optical/UV emitting region. The corresponding photon energy density $u_{\rm opt/UV} \approx L_{\rm opt/UV}/\big( 4 \pi R_{\rm sh}^2 c \big) \sim 2.7 \times 10^{-4}\ {\rm erg\ cm^{-3}} (L_{\rm opt/UV}/10^{44}\ {\rm erg\ s^{-1}}) (R_{\rm sh}/10^{18}\ {\rm cm})^{-2}$. Thus, the photon energy density is reduced by a factor of $(R_{\rm em}/R_{\rm sh})^2$ relative to that near an optical/UV emitting region of characteristic scale $R_{\rm em}$ (for typical $R_{\rm em} \sim 10^{15}\ {\rm cm}$, this reduction is $\sim 10^{-6}$). Thus, the combination of the relatively high proton energy needed for the $p\gamma$ ($\Delta$-resonance) threshold and the geometric dilution of the direct optical/UV radiation field suppresses the $p\gamma$ contribution relative to the $pp$ channel considered here. In the obscured IR-only TDE scenario, a substantial fraction of the primary optical/UV radiation is additionally reprocessed by the surrounding dust. We therefore expect neutrino production through interactions with the direct optical/UV radiation field to be subdominant to the $pp$ contribution.
\subsubsection{Post-shock free-free photons}
An additional target photon field can arise locally from thermal free-free emission by the shocked gas~\citep{1986rpa..book.....R}. Once the shock becomes radiative, a substantial fraction of the kinetic power dissipated at the shock can be converted into radiation, with the thermal bremsstrahlung spectrum extending up to photon energies of order $k_BT_{\rm ds}$~\citep{2012ApJ...747L..17C,2013ApJ...763...42O}. Unlike the central optical/UV radiation, these photons are produced locally in the shocked region and therefore
do not suffer from geometric dilution between the central source and the forward shock. We conservatively take $L_{\rm ff}\simeq L_{\rm sh}$, corresponding to $u_{\rm ff}\simeq L_{\rm sh}/(4\pi R_{\rm sh}^2c)$, and approximate the free-free spectrum as $L_{\varepsilon_\gamma}\propto
\exp(-\varepsilon_\gamma/k_BT_{\rm ds})$. At the onset of the radiative phase in our IR-only model, $t_{\rm crit} \simeq 9.8\ {\rm yrs}$, we find $k_B T_{\rm ds} \simeq 12\ {\rm keV}$ and $\varepsilon_p^{\rm max} \simeq 1.8 \times 10^7\ {\rm GeV}$, corresponding to a characteristic target photon energy $\varepsilon_{\gamma} \simeq 17\ {\rm eV}$ for the $\Delta$-resonance. Thus, unlike the dust-reprocessed IR photons, photomeson production on the free-free radiation is kinematically allowed for the highest-energy protons. Nevertheless, we find $u_{\rm ff} \simeq 9 \times 10^{-4}\ {\rm erg\ cm^{-3}}$ and $n_{\gamma,\rm ff} \sim 3 \times 10^5\ {\rm cm^{-3}}$, corresponding to $t_{p\gamma}^{\rm ff} \sim 10^{12}\ {\rm s}$ for $\sigma_{p\gamma} \kappa_{p\gamma} \sim 10^{-28}\ {\rm cm^2}$. This is much longer than the $pp$ interaction time even before accounting for the enhanced target density due to radiative compression. We therefore conclude that $p\gamma$ interactions with the post-shock free-free radiation are also subdominant.
\subsection{Caveats}
\label{subsec:caveats}
In this subsection, we briefly discuss several caveats and assumptions we make in the current work. We leave a detailed analysis to address these caveats for future work. First, the CNM profile assumed for the IR-only TDE population should be regarded as an optimistic benchmark rather than a uniquely determined description of these systems. The observed IR echoes primarily constrain the characteristic location of the reprocessing dust, while the absence of an optical counterpart constrains the obscuring column along the line of sight. Neither measurement uniquely determines the radial gas-density profile. Different density distributions, geometries, and covering factors can therefore satisfy these observational constraints. In particular, our choice of $n_{\rm env} = 10^6\ {\rm cm^{-3}}$ represents one realization of a dense circumnuclear environment and the predicted neutrino yield depends sensitively on this assumption.

Second, our treatment of the radiative shock is necessarily idealized. Once $t_{\rm ff}<t_{\rm dyn}$, we approximate the subsequent bulk evolution as a momentum-conserving shell, matched continuously onto the preceding energy-conserving solution at $t_{\rm crit}$. This instantaneous transition neglects the detailed structure of the cooling region and any intermediate pressure-driven phase. Similarly, the density enhancement of the cooled shocked material is estimated assuming approximately isobaric cooling to $T_c = 10^4$ K. A hydrodynamical treatment including radiative cooling would be required to determine the shock evolution and compression more self-consistently. Our simplified one-zone treatment assumes that CR protons accelerated near the forward shock remain in the shocked downstream material as it cools and compresses, allowing them to efficiently interact with the radiatively enhanced density. However, the extent to which the CRs penetrate and remain confined within this dense cooling layer depends on their transport through the shocked region, which is not spatially resolved in our one-zone calculation. The resulting $n_{\rm cool}$ should therefore be interpreted as an effective, maximally compressed target density rather than a detailed prediction for the density sampled by the CRs. In the future, a more detailed multi-zone treatment including CR transport and interactions in the downstream region would be required to quantify this effect more self-consistently.

An additional uncertainty concerns the spatial extent of the particle acceleration region once the shock becomes radiative. As discussed in Section~\ref{subsec:cr_acc}, our $t_{\rm acc} < t_{\rm ff}$ prescription approximately corresponds to requiring particle confinement within the laminar radiative cooling layer. Multidimensional radiative shocks, however, can develop thin-shell instabilities and turbulent mixing between the hot post-shock gas and the cold dense shell~\citep{Steinberg:2018pas,Diesing:2026uog}. Such mixing may reduce the effective thickness of the hot acceleration region below the laminar cooling length. The magnitude of this additional suppression is uncertain for TDE outflow–CNM shocks and is not included in our one-zone treatment. If efficient, it would further reduce $\varepsilon_p^{\rm max}$ during the radiative phase, suppressing the neutrino emission above the corresponding energy cut-off.

A related uncertainty is the ionization state of the upstream CNM. In a partially ionized medium, ion-neutral interactions can modify the shock structure and CR spectrum, while ion-neutral damping can suppress the magnetic turbulence required for efficient particle acceleration~\citep{Drury:1995ep,Reville:2007er,Blasi:2012xe,Metzger:2015zka}. Radiation from the shocked gas may instead maintain a photoionized layer immediately upstream, but determining its extent and ionization state requires a detailed photoionization calculation beyond the scope of this work. We neglect these effects, which, if important, could further reduce $\varepsilon_p^{\rm max}$ and suppress the high-energy neutrino emission.

Finally, the population predictions depend on the uncertain occurrence rates and outflow properties of TDEs. Our choice of $\dot{\rho}_{\rm IR,0} = \dot{\rho}_{\rm opt,0}/3$ is an approximate population level prediction motivated by the currently inferred rates, rather than a direct measurement of the ratio between the IR-only and optical TDEs. Furthermore, we assume that each population launch outflows whose interaction with the CNM can be represented by the corresponding model adopted above. If only a fraction $f_{\rm out}$ of either population produces such outflows, its diffuse neutrino contribution scales approximately linearly with $f_{\rm out}$. The normalization of the diffuse intensity likewise scales directly with the uncertain local occurrence rates, while its cosmological contribution additionally depends on the poorly constrained redshift evolution of the TDE population.
\subsection{Comparison with previous work}
Several scenarios for high-energy neutrino production in TDEs have been proposed. The outflow-cloud interaction model where outflows produced by TDEs interact with dense clouds seen as broad-line regions of AGN was discussed in~\citet{Wu:2021vaw}. The number density in the cloud is so high that all the accelerated cosmic rays can experience hadronuclear interactions in this model. However, it is unclear whether such dense clouds exist in TDE environments, because TDEs usually occur in quiescent galaxies, which do not have broad-line regions.

In~\cite{Winter:2022fpf,Yuan:2024foi} neutrino production by photohadronic interactions was discussed. They considered compact dissipation regions close to the photosphere that produce optical/UV emissions. This leads to a high photohadronic interaction rate compared to our model, although the dissipation mechanism that enables cosmic ray acceleration at the vicinity of photosphere is not specified. 

Furthermore,~\citet{Murase:2020lnu} considered a few scenarios, including neutrino production in hot coronae, hot accretion flows, and hidden winds. For accretion flow models, it is under debate what fraction of the disrupted stellar material falls to the SMBH, and thus, the energy budget of accretion flow models is uncertain. For the hidden wind model, protons are accelerated inside the quasi-spherical TDE debris surrounding the SMBH. The accelerated protons efficiently interact with the debris via hadronuclear interactions, leading to efficient neutrino production. This scenario is feasible as long as cosmic-ray protons are accelerated inside debris.

Our scenario considers outflow-CNM interactions, which lead to lower neutrino production efficiencies because of much larger dissipation radii than the models above. Nevertheless, our setup is supported by radio observations, and thus, our model can be regarded as one of the most conservative neutrino production scenarios in the TDE environment.
\section{Conclusion}
\label{sec:concl}
%
TDEs can launch sub-relativistic outflows (or winds) that drive shocks into the CNM surrounding the central SMBH. Such scenarios provide a natural site for CR acceleration and eventual high-energy neutrino production through hadronuclear ($pp$) interactions (see Figure~\ref{fig:schematic}). In this work, we have investigated high-energy neutrino emission from TDE outflow-CNM interactions, considering two distinct populations - optical TDEs and IR-only TDEs, the latter of which can occur in substantially denser and more obscured environments (see Figure~\ref{fig:density}). We model the evolution of the forward shock (see Figure~\ref{fig:rsh_vsh}), acceleration and quasi-steady state transport of the CR protons, and their subsequent $pp$ interactions with the shocked CNM. For the IR-only TDEs in dense environments, we additionally account for the transition to a radiative shock, including the resulting shock deceleration and the enhanced compression of the cooled downstream material (see Section~\ref{subsec:cool_shock}).

For the optical TDEs, the shocked material remains effectively adiabatic and the source remains optically thin to $pp$ interactions, resulting in a comparatively small neutrino yield (see Figures~\ref{fig:luminosity} and~\ref{fig:nu_spectra}). For the IR-only TDEs, the $pp$ interaction efficiency can instead approach unity, making the source nearly calorimetric. At the same time, the enhanced swept-up mass (Figure~\ref{fig:mass_swept}) and the eventual radiative evolution cause stronger shock deceleration and reduce the maximum energy to which protons can be accelerated at late times (Figure~\ref{fig:epmax_fpp}). The dense environment therefore enhances the efficiency of neutrino production while shifting the high-energy cutoff to lower energies. For a cosmological population, we find that optically-bright TDEs contribute only a negligible fraction of the observed diffuse neutrino flux, whereas the IR-only population can contribute approximately $\sim 1 - 5\%$ at $100\ {\rm TeV}$ and can approach $\sim 10\%$, considering rate uncertainties and a more strongly evolving population scenario (see Figure~\ref{fig:diffuse}).

The non-detectability of a single source and the long-lasting neutrino emission (see Figure~\ref{fig:nu_spectra}), also motivate the targeted stacking searches using EM observations and catalogs of TDEs. We explored the prospects of such searches using track-like events in IceCube, IceCube-Gen2, KM3NeT, and HUNT, consistently accounting for the competing effects of increasing source statistics, source distance, search time window, and neutrino backgrounds. For the IR-only TDEs, HUNT can reach $3\sigma$ C.L. sensitivity for relatively nearby ($z \lesssim 0.4$) TDEs and multi-year search windows (see Figure~\ref{fig:hunt_sig}). The rapidly growing TDE samples expected from wide-field optical and infrared surveys, including Rubin/LSST, WFST, Roman, and SPHEREx, will therefore provide increasingly powerful catalogs for such searches (see Section~\ref{subsec:em_surveys}). 

In particular, complementary optical and IR-selected samples may enable population-resolved neutrino searches capable of testing whether the dense circumnuclear environments associated with obscured TDEs lead to enhanced high-energy neutrino production. Unlike neutrino production scenarios associated with relativistic TDE jets, the outflow–CNM interaction considered here does not require a rare or favorably oriented jet. Non-relativistic outflows interacting with the surrounding nuclear environment therefore provide a concrete target for high-energy neutrino searches, motivating dedicated searches for this emission in current and future data from neutrino observatories.
\begin{acknowledgements}
We thank Damiano Caprioli, Brian D. Metzger, and Walter Winter for useful feedback on the manuscript. M. \,M. wishes to thank the Astronomical Institute at Tohoku University for their hospitality where a major part
of this work was completed. M.\,M. acknowledges support from the FermiForward Discovery Group, LLC under Contract No. 89243024CSC000002 with the U.S. Department of Energy, Office of Science, Office of High Energy Physics. M.\,M. also acknowledges the support of the NSF-Simons AI-Institute for the Sky (SkAI) via grants NSF AST-2421845 and Simons Foundation MPS-AI-00010513. S.\,S.\,K acknowledges the support by KAKENHI No. 23H04899, 26K00733, 26K00696, and the Tohoku Initiative for Fostering Global Researchers for Interdisciplinary Sciences (TI-FRIS) of MEXT’s Strategic Professional Development Program for Young Researchers. T.\,M. acknowledges the support by KAKENHI No. 23H04900 and 24K17088.
\end{acknowledgements}
\bibliography{refs}{}

\begin{thebibliography}{}
\expandafter\ifx\csname natexlab\endcsname\relax\def\natexlab#1{#1}\fi
\providecommand{\url}[1]{\href{#1}{#1}}
\providecommand{\dodoi}[1]{doi:~\href{http://doi.org/#1}{\nolinkurl{#1}}}
\providecommand{\doeprint}[1]{\href{http://ascl.net/#1}{\nolinkurl{http://ascl.net/#1}}}
\providecommand{\doarXiv}[1]{\href{https://arxiv.org/abs/#1}{\nolinkurl{https://arxiv.org/abs/#1}}}

\bibitem[{M.~G. Aartsen {et~al.}(2018)Aartsen {et~al.}}]{IceCube:2018fhm}
Aartsen, M.~G., {et~al.} 2018, \bibinfo{title}{{Differential limit on the extremely-high-energy cosmic neutrino flux in the presence of astrophysical background from nine years of IceCube data},} Phys. Rev. D, 98, 062003, \dodoi{10.1103/PhysRevD.98.062003}

\bibitem[{R. Abbasi {et~al.}(2025)Abbasi {et~al.}}]{IceCube:2025uzh}
Abbasi, R., {et~al.} 2025, \bibinfo{title}{{IceCat-2: Updated IceCube Event Catalog of Alert Tracks},} PoS, ICRC2025, 1224, \dodoi{10.22323/1.501.1224}

\bibitem[{R. Abbasi {et~al.}(2026)Abbasi {et~al.}}]{Abbasi:2026ehs}
Abbasi, R., {et~al.} 2026, \bibinfo{title}{{IceCube Second Track Data Release IceTracks-DR2: Data from 2008-2022 for Neutrino Source Searches},} \doarXiv{2605.19040}

\bibitem[{R. {Abbasi} {et~al.}(2026){Abbasi}, {Ackermann}, {Adams}, {Aguilar}, {Ahlers}, {Alameddine}, {Ali}, {Amin}, {Andeen}, {Arg\{{\"u}\}elles}, {Athanasiadou}, {Axani}, {Babu}, {Bai}, {Balagopal V.}, {Barwick}, {Basu}, {Bay}, {Beatty}, {Becker Tjus}, {Behrens}, {Beise}, {Bellenghi}, {Benkel}, {BenZvi}, {Berley}, {Bernardini}, {Besson}, {Blaufuss}, {Bloom}, {Blot}, {Bontempo}, {Book Motzkin}, {Boscolo Meneguolo}, {B\{{\"o}\}ser}, {Botner}, {B\{{\"o}\}ttcher}, {Braun}, {Brinson}, {Brisson-Tsavoussis}, {Brusa}, {Burley}, {Butterfield}, {Carloni}, {Carpio}, {Chau}, {Chen}, {Chen}, {Chirkin}, {Choi}, {Chubarov}, {Clark}, {Coloma Borja}, {Connolly}, {Conrad}, {Cowen}, {De Clercq}, {DeLaunay}, {Delgado}, {Delmeulle}, {Deng}, {Desiati}, {de Vries}, {de Wasseige}, {DeYoung}, {D\{\textbackslash'{\i}\}az-V\{{\'e}\}lez}, {DiKerby}, {Ding}, {Dittmer}, {Domi}, {Draper}, {Dueser}, {Durnford}, {Dutta}, {DuVernois}, {Ehrhardt}, {Eidenschink}, {Eimer}, {Eldridge}, {Eller}, {Ellinger}, {Els\{{\"a}\}sser}, {Engel}, {Erpenbeck}, {Esmail}, {Eulig}, {Evans}, {Evenson}, {Fan}, {Fang}, {Farrag}, {Fattorini}, {Fazely}, {Fedynitch}, {Feigl}, {Finley}, {Fox}, {Franckowiak}, {Fukami}, {F\{{\"u}\}rst}, {Gallagher}, {Ganster}, {Garcia}, {Garcia}, {Genton}, {Gerhardt}, {Ghadimi}, {Glaser}, {Gl\{{\"u}\}senkamp}, {Gonzalez}, {Goswami}, {Granados}, {Grant}, {Gray}, {Griffin}, {Griswold}, {Groth}, {Guevel}, {G\{{\"u}\}nther}, {Gutjahr}, {Ha}, {Hallgren}, {Halve}, {Halzen}, {Hamacher}, {Handt}, {Hanson}, {Hardin}, {Harnisch}, {Hatch}, {Haungs}, {H\{{\"a}\}u{\ss}ler}, {Helbing}, {Hellrung}, {Henke}, {Hennig}, {Henningsen}, {Heuermann}, {Hewett}, {Heyer}, {Hickford}, {Hidvegi}, {Hill}, {Hill}, {Hmaid}, {Hoffman}, {Hollnagel}, {Hooper}, {Hori}, {Hoshina}, {Hostert}, {Hou}, {Hrywniak}, {Huber}, {Hultqvist}, {Hymon}, {Ishihara}, {Iwakiri}, {Jacquart}, {Jain}, {Janik}, {Jansson}, {Jin}, {Kamp}, {Kang}, {Kang}, {Kappes}, {Kardum}, {Karg}, {Karle}, {Katil}, {Kauer}, {Kelley}, {Khanal}, {Khatee Zathul}, {Kheirandish}, {Kim}, {Kimku}, {Kirchner}, {Kiryluk}, {Klein}, {Klein}, {Kobayashi}, {Koch}, {Kochocki}, {Koirala}, {Kolanoski}, {Kontrimas}, {K\{{\"o}\}pke}, {Kopper}, {Koskinen}, {Koundal}, {Kowalski}, {Kozynets}, {Kravka}, {Krieger}, {Krishnan}, {Kruiswijk}, {Krupczak}, \& {Kun}}]{2026arXiv260900657A}
{Abbasi}, R., {Ackermann}, M., {Adams}, J., {et~al.} 2026, \bibinfo{title}{{Search for Neutrinos from Tidal Disruption Events with IceCube},} arXiv e-prints, arXiv:2609.00657, \dodoi{10.48550/arXiv.2609.00657}

\bibitem[{R. Abbasi {et~al.}(2026)Abbasi {et~al.}}]{IceCube:2025tgp}
Abbasi, R., {et~al.} 2026, \bibinfo{title}{{Evidence for a Spectral Break or Curvature in the Spectrum of Astrophysical Neutrinos from 5 TeV--10 PeV},} Phys. Rev. Lett., 136, 121002, \dodoi{10.1103/2gh9-d4q7}

\bibitem[{A. Abdul~Halim {et~al.}(2023)Abdul~Halim {et~al.}}]{PierreAuger:2023pjg}
Abdul~Halim, A., {et~al.} 2023, \bibinfo{title}{{Latest results from the searches for ultra-high-energy photons and neutrinos at the Pierre Auger Observatory},} PoS, ICRC2023, 1488, \dodoi{10.22323/1.444.1488}

\bibitem[{S. Adrian-Martinez {et~al.}(2016)Adrian-Martinez {et~al.}}]{KM3Net:2016zxf}
Adrian-Martinez, S., {et~al.} 2016, \bibinfo{title}{{Letter of intent for KM3NeT 2.0},} J. Phys. G, 43, 084001, \dodoi{10.1088/0954-3899/43/8/084001}

\bibitem[{A. Albert {et~al.}(2024)Albert {et~al.}}]{ANTARES:2024ihw}
Albert, A., {et~al.} 2024, \bibinfo{title}{{Constraints on the energy spectrum of the diffuse cosmic neutrino flux from the ANTARES neutrino telescope},} JCAP, 08, 038, \dodoi{10.1088/1475-7516/2024/08/038}

\bibitem[{K.~D. Alexander {et~al.}(2020)Alexander, van Velzen, Horesh, \& Zauderer}]{Alexander:2020xwb}
Alexander, K.~D., van Velzen, S., Horesh, A., \& Zauderer, B.~A. 2020, \bibinfo{title}{{Radio Properties of Tidal Disruption Events},} Space Sci. Rev., 216, 81, \dodoi{10.1007/s11214-020-00702-w}

\bibitem[{P. Blasi {et~al.}(2012)Blasi, Morlino, Bandiera, Amato, \& Caprioli}]{Blasi:2012xe}
Blasi, P., Morlino, G., Bandiera, R., Amato, E., \& Caprioli, D. 2012, \bibinfo{title}{{Collisionless shocks in a partially ionized medium: I. Neutral return flux and its effects on acceleration of test particles},} Astrophys. J., 755, 121, \dodoi{10.1088/0004-637X/755/2/121}

\bibitem[{D.~N. Burrows {et~al.}(2011)Burrows {et~al.}}]{Burrows:2011dn}
Burrows, D.~N., {et~al.} 2011, \bibinfo{title}{{Discovery of the Onset of Rapid Accretion by a Dormant Massive Black Hole},} Nature, 476, 421, \dodoi{10.1038/nature10374}

\bibitem[{Y. Cendes {et~al.}(2024)Cendes {et~al.}}]{Cendes:2023rrc}
Cendes, Y., {et~al.} 2024, \bibinfo{title}{{Ubiquitous Late Radio Emission from Tidal Disruption Events},} Astrophys. J., 971, 185, \dodoi{10.3847/1538-4357/ad5541}

\bibitem[{M. Chen(2026)Chen}]{Chen:2026jdx}
Chen, M. 2026, \bibinfo{title}{{HUNT: An ultra-large-scale neutrino astronomy telescope},} Nucl. Instrum. Meth. A, 1086, 171374, \dodoi{10.1016/j.nima.2026.171374}

\bibitem[{R.~A. {Chevalier} \& C.~M. {Irwin}(2012){Chevalier} \& {Irwin}}]{2012ApJ...747L..17C}
{Chevalier}, R.~A., \& {Irwin}, C.~M. 2012, \bibinfo{title}{{X-Rays from Supernova Shocks in Dense Mass Loss},} \apjl, 747, L17, \dodoi{10.1088/2041-8205/747/1/L17}

\bibitem[{B.~P. {Crill} {et~al.}(2020){Crill}, {Werner}, {Akeson}, {Ashby}, {Bleem}, {Bock}, {Bryan}, {Burnham}, {Byunh}, {Chang}, {Chiang}, {Cook}, {Cooray}, {Davis}, {Dor{\'e}}, {Dowell}, {Dubois-Felsmann}, {Eifler}, {Faisst}, {Habib}, {Heinrich}, {Heitmann}, {Heaton}, {Hirata}, {Hristov}, {Hui}, {Jeong}, {Kang}, {Kecman}, {Kirkpatrick}, {Korngut}, {Krause}, {Lee}, {Lisse}, {Masters}, {Mauskopf}, {Melnick}, {Miyasaka}, {Nayyeri}, {Nguyen}, {{\"O}berg}, {Padin}, {Paladini}, {Pourrahmani}, {Pyo}, {Smith}, {Song}, {Symons}, {Teplitz}, {Tolls}, {Unwin}, {Windhorst}, {Yang}, \& {Zemcov}}]{2020SPIE11443E..0IC}
{Crill}, B.~P., {Werner}, M., {Akeson}, R., {et~al.} 2020, \bibinfo{title}{{SPHEREx: NASA's near-infrared spectrophotometric all-sky survey},} in Society of Photo-Optical Instrumentation Engineers (SPIE) Conference Series, Vol. 11443, Space Telescopes and Instrumentation 2020: Optical, Infrared, and Millimeter Wave, ed. M.~{Lystrup} \& M.~D. {Perrin}, 114430I, \dodoi{10.1117/12.2567224}

\bibitem[{L. Dai \& K. Fang(2017)Dai \& Fang}]{Dai:2016gtz}
Dai, L., \& Fang, K. 2017, \bibinfo{title}{{Can tidal disruption events produce the IceCube neutrinos?},} Mon. Not. Roy. Astron. Soc., 469, 1354, \dodoi{10.1093/mnras/stx863}

\bibitem[{F. De~Colle \& W. Lu(2020)De~Colle \& Lu}]{DeColle:2019wzp}
De~Colle, F., \& Lu, W. 2020, \bibinfo{title}{{Jets from Tidal Disruption Events},} New Astron. Rev., 89, 101538, \dodoi{10.1016/j.newar.2020.101538}

\bibitem[{R. Diesing \& B.~D. Metzger(2026)Diesing \& Metzger}]{Diesing:2026uog}
Diesing, R., \& Metzger, B.~D. 2026, \bibinfo{title}{{A Unified Model for Shock Interaction and {\ensuremath{\gamma}}-Ray Emission in Classical Novae},} Astrophys. J., 1004, 109, \dodoi{10.3847/1538-4357/ae6db1}

\bibitem[{L.~O. {Drury}(1983){Drury}}]{1983RPPh...46..973D}
{Drury}, L.~O. 1983, \bibinfo{title}{{REVIEW ARTICLE: An introduction to the theory of diffusive shock acceleration of energetic particles in tenuous plasmas},} Reports on Progress in Physics, 46, 973, \dodoi{10.1088/0034-4885/46/8/002}

\bibitem[{L.~O. Drury {et~al.}(1996)Drury, Duffy, \& Kirk}]{Drury:1995ep}
Drury, L.~O., Duffy, P., \& Kirk, J.~G. 1996, \bibinfo{title}{{Limits on diffusive shock acceleration in dense and incompletely ionised media},} Astron. Astrophys., 309, 1002.
\newblock \doarXiv{astro-ph/9510066}

\bibitem[{S. Gezari(2021)Gezari}]{Gezari:2021bmb}
Gezari, S. 2021, \bibinfo{title}{{Tidal Disruption Events},} Ann. Rev. Astron. Astrophys., 59, 21, \dodoi{10.1146/annurev-astro-111720-030029}

\bibitem[{K. Hayasaki \& R. Yamazaki(2019)Hayasaki \& Yamazaki}]{Hayasaki:2019kjy}
Hayasaki, K., \& Yamazaki, R. 2019, \bibinfo{title}{{Neutrino Emissions from Tidal Disruption Remnants},} \dodoi{10.3847/1538-4357/ab44ca}

\bibitem[{A. Horesh {et~al.}(2021)Horesh, Cenko, \& Arcavi}]{Horesh:2021gvp}
Horesh, A., Cenko, S.~B., \& Arcavi, I. 2021, \bibinfo{title}{{Delayed Radio Flares from a Tidal Disruption Event},} Nature Astron., 5, 491, \dodoi{10.1038/s41550-021-01300-8}

\bibitem[{N. Jiang {et~al.}(2023)Jiang, Zhou, Zhu, Wang, \& Wang}]{Jiang:2023kbb}
Jiang, N., Zhou, Z., Zhu, J., Wang, Y., \& Wang, T. 2023, \bibinfo{title}{{Two Candidate Obscured Tidal Disruption Events Coincident with High-energy Neutrinos},} Astrophys. J. Lett., 953, L12, \dodoi{10.3847/2041-8213/acebe3}

\bibitem[{M. Karmen {et~al.}(2026)Karmen, Gezari, Norman, \& Guolo}]{Karmen:2026ixp}
Karmen, M., Gezari, S., Norman, C., \& Guolo, M. 2026, \bibinfo{title}{{Tidal Disruption Event Rates across Cosmic Time: Forecasts for LSST, Roman, and JWST and Their Constraints on the Supermassive Black Hole Mass Function},} Astrophys. J., 1006, 20, \dodoi{10.3847/1538-4357/ae7a49}

\bibitem[{K. Kashiyama \& P. Meszaros(2014)Kashiyama \& Meszaros}]{Kashiyama:2014rza}
Kashiyama, K., \& Meszaros, P. 2014, \bibinfo{title}{{Galaxy Mergers as a Source of Cosmic Rays, Neutrinos, and Gamma Rays},} Astrophys. J. Lett., 790, L14, \dodoi{10.1088/2041-8205/790/1/L14}

\bibitem[{S.~S. Kimura \& T.~J. Moriya(2025)Kimura \& Moriya}]{Kimura:2024lvt}
Kimura, S.~S., \& Moriya, T.~J. 2025, \bibinfo{title}{{High-energy Gamma-Ray and Neutrino Emissions from Interacting Supernovae Based on Radiation Hydrodynamic Simulations: A Case of SN 2023ixf},} Astrophys. J., 984, 103, \dodoi{10.3847/1538-4357/adc716}

\bibitem[{S. Komossa(2015)Komossa}]{Komossa:2015qya}
Komossa, S. 2015, \bibinfo{title}{{Tidal disruption of stars by supermassive black holes: Status of observations},} JHEAp, 7, 148, \dodoi{10.1016/j.jheap.2015.04.006}

\bibitem[{Z. Lin {et~al.}(2022)Lin, Jiang, \& Kong}]{Lin:2022npv}
Lin, Z., Jiang, N., \& Kong, X. 2022, \bibinfo{title}{{The prospects of finding tidal disruption events with 2.5-m Wide-Field Survey Telescope based on mock observations},} Mon. Not. Roy. Astron. Soc., 513, 2422, \dodoi{10.1093/mnras/stac946}

\bibitem[{R.-Y. Liu {et~al.}(2020)Liu, Xi, \& Wang}]{Liu:2020isi}
Liu, R.-Y., Xi, S.-Q., \& Wang, X.-Y. 2020, \bibinfo{title}{{Neutrino emission from an off-axis jet driven by the tidal disruption event AT2019dsg},} Phys. Rev. D, 102, 083028, \dodoi{10.1103/PhysRevD.102.083028}

\bibitem[{M.~S. {Longair}(2011){Longair}}]{2011hea..book.....L}
{Longair}, M.~S. 2011, {High Energy Astrophysics}

\bibitem[{W. Lu {et~al.}(2016)Lu, Kumar, \& Evans}]{Lu:2015jfg}
Lu, W., Kumar, P., \& Evans, N.~J. 2016, \bibinfo{title}{{Infrared emission from tidal disruption events {\textendash} probing the pc-scale dust content around galactic nuclei},} Mon. Not. Roy. Astron. Soc., 458, 575, \dodoi{10.1093/mnras/stw307}

\bibitem[{C. Lunardini \& W. Winter(2017)Lunardini \& Winter}]{Lunardini:2016xwi}
Lunardini, C., \& Winter, W. 2017, \bibinfo{title}{{High Energy Neutrinos from the Tidal Disruption of Stars},} Phys. Rev. D, 95, 123001, \dodoi{10.1103/PhysRevD.95.123001}

\bibitem[{M. Masterson {et~al.}(2024)Masterson {et~al.}}]{Masterson:2024rur}
Masterson, M., {et~al.} 2024, \bibinfo{title}{{A New Population of Mid-infrared-selected Tidal Disruption Events: Implications for Tidal Disruption Event Rates and Host Galaxy Properties},} Astrophys. J., 961, 211, \dodoi{10.3847/1538-4357/ad18bb}

\bibitem[{R. Matsui {et~al.}(2023)Matsui, Kimura, Toma, \& Murase}]{Matsui:2023ohr}
Matsui, R., Kimura, S.~S., Toma, K., \& Murase, K. 2023, \bibinfo{title}{{High-energy Neutrino Emission Associated with Gravitational-wave Signals: Effects of Cocoon Photons and Constraints on Late-time Emission},} Astrophys. J., 950, 190, \dodoi{10.3847/1538-4357/acd004}

\bibitem[{T. Matsumoto \& T. Piran(2021)Matsumoto \& Piran}]{Matsumoto:2021gcg}
Matsumoto, T., \& Piran, T. 2021, \bibinfo{title}{{Radio constraint on outflows from tidal disruption events},} Mon. Not. Roy. Astron. Soc., 507, 4196, \dodoi{10.1093/mnras/stab2418}

\bibitem[{T. Matsumoto \& T. Piran(2024)Matsumoto \& Piran}]{Matsumoto:2024eja}
Matsumoto, T., \& Piran, T. 2024, \bibinfo{title}{{Late-time Radio Flares in Tidal Disruption Events},} Astrophys. J., 971, 49, \dodoi{10.3847/1538-4357/ad58ba}

\bibitem[{B.~D. Metzger {et~al.}(2016)Metzger, Caprioli, Vurm, Beloborodov, Bartos, \& Vlasov}]{Metzger:2015zka}
Metzger, B.~D., Caprioli, D., Vurm, I., {et~al.} 2016, \bibinfo{title}{{Novae as Tevatrons: Prospects for CTA and IceCube},} Mon. Not. Roy. Astron. Soc., 457, 1786, \dodoi{10.1093/mnras/stw123}

\bibitem[{M. Mukhopadhyay {et~al.}(2024{\natexlab{a}})Mukhopadhyay, Bhattacharya, \& Murase}]{Mukhopadhyay:2023mld}
Mukhopadhyay, M., Bhattacharya, M., \& Murase, K. 2024{\natexlab{a}}, \bibinfo{title}{{Multimessenger signatures of delayed choked jets in tidal disruption events},} Mon. Not. Roy. Astron. Soc., 534, 1528, \dodoi{10.1093/mnras/stae2080}

\bibitem[{M. Mukhopadhyay {et~al.}(2024{\natexlab{b}})Mukhopadhyay, Kimura, \& Murase}]{Mukhopadhyay:2023niv}
Mukhopadhyay, M., Kimura, S.~S., \& Murase, K. 2024{\natexlab{b}}, \bibinfo{title}{{Gravitational wave triggered searches for high-energy neutrinos from binary neutron star mergers: Prospects for next generation detectors},} Phys. Rev. D, 109, 043053, \dodoi{10.1103/PhysRevD.109.043053}

\bibitem[{M. Mukhopadhyay {et~al.}(2026{\natexlab{a}})Mukhopadhyay, Kimura, Vurm, \& Metzger}]{Mukhopadhyay:2026afk}
Mukhopadhyay, M., Kimura, S.~S., Vurm, I., \& Metzger, B.~D. 2026{\natexlab{a}}, \bibinfo{title}{{Neutrino and Electromagnetic Signatures from Superluminous Supernovae: A Case Study for SN 2017egm},} Astrophys. J., 1006, 24, \dodoi{10.3847/1538-4357/ae7d21}

\bibitem[{M. Mukhopadhyay {et~al.}(2024{\natexlab{c}})Mukhopadhyay, Kotera, Wissel, Murase, \& Kimura}]{Mukhopadhyay:2024lwq}
Mukhopadhyay, M., Kotera, K., Wissel, S., Murase, K., \& Kimura, S.~S. 2024{\natexlab{c}}, \bibinfo{title}{{Ultrahigh-energy neutrino searches using next-generation gravitational wave detectors at radio neutrino detectors: GRAND, IceCube-Gen2 Radio, and RNO-G},} Phys. Rev. D, 110, 063004, \dodoi{10.1103/PhysRevD.110.063004}

\bibitem[{M. Mukhopadhyay {et~al.}(2026{\natexlab{b}})Mukhopadhyay, Wusinich, \& Murase}]{Mukhopadhyay:2026qlq}
Mukhopadhyay, M., Wusinich, P., \& Murase, K. 2026{\natexlab{b}}, \bibinfo{title}{{Constraining high-energy neutrinos from tidal disruption events with IceCube high-energy starting events},} \doarXiv{2601.20934}

\bibitem[{K. Murase {et~al.}(2020)Murase, Kimura, Zhang, Oikonomou, \& Petropoulou}]{Murase:2020lnu}
Murase, K., Kimura, S.~S., Zhang, B.~T., Oikonomou, F., \& Petropoulou, M. 2020, \bibinfo{title}{{High-Energy Neutrino and Gamma-Ray Emission from Tidal Disruption Events},} Astrophys. J., 902, 108, \dodoi{10.3847/1538-4357/abb3c0}

\bibitem[{P. Nair {et~al.}(2026)Nair, Panagiotou, Masterson, De, Kara, Winkler, \& Rdan}]{Nair:2026lfv}
Nair, P., Panagiotou, C., Masterson, M., {et~al.} 2026, \bibinfo{title}{{A Suppressed Volumetric Rate of High-luminosity Mid-infrared Selected Tidal Disruption Events},} Astrophys. J., 1007, 40, \dodoi{10.3847/1538-4357/ae83a1}

\bibitem[{J. Necker {et~al.}(2025)Necker, Graikou, Kowalski, Franckowiak, Nordin, Pernice, van Velzen, \& Veres}]{Necker:2024vfs}
Necker, J., Graikou, E., Kowalski, M., {et~al.} 2025, \bibinfo{title}{{Flaires - A comprehensive catalog of dust echo-like infrared flares},} Astron. Astrophys., 695, A228, \dodoi{10.1051/0004-6361/202451340}

\bibitem[{E.~O. {Ofek} {et~al.}(2013){Ofek}, {Fox}, {Cenko}, {Sullivan}, {Gnat}, {Frail}, {Horesh}, {Corsi}, {Quimby}, {Gehrels}, {Kulkarni}, {Gal-Yam}, {Nugent}, {Yaron}, {Filippenko}, {Kasliwal}, {Bildsten}, {Bloom}, {Poznanski}, {Arcavi}, {Laher}, {Levitan}, {Sesar}, \& {Surace}}]{2013ApJ...763...42O}
{Ofek}, E.~O., {Fox}, D., {Cenko}, S.~B., {et~al.} 2013, \bibinfo{title}{{X-Ray Emission from Supernovae in Dense Circumstellar Matter Environments: A Search for Collisionless Shocks},} \apj, 763, 42, \dodoi{10.1088/0004-637X/763/1/42}

\bibitem[{M.~J. Rees(1988)Rees}]{Rees1988}
Rees, M.~J. 1988, \bibinfo{title}{Tidal disruption of stars by black holes of $10^{6}$--$10^{8}$ solar masses in nearby galaxies,} Nature, 333, 523, \dodoi{10.1038/333523a0}

\bibitem[{S. Reusch {et~al.}(2022)Reusch {et~al.}}]{Reusch:2021ztx}
Reusch, S., {et~al.} 2022, \bibinfo{title}{{Candidate Tidal Disruption Event AT2019fdr Coincident with a High-Energy Neutrino},} Phys. Rev. Lett., 128, 221101, \dodoi{10.1103/PhysRevLett.128.221101}

\bibitem[{B. Reville {et~al.}(2007)Reville, Kirk, Duffy, \& Sullivan}]{Reville:2007er}
Reville, B., Kirk, J.~G., Duffy, P., \& Sullivan, S.~O. 2007, \bibinfo{title}{{A cosmic ray current driven instability in partially ionised media},} Astron. Astrophys., 475, 435, \dodoi{10.1051/0004-6361:20078336}

\bibitem[{G.~B. {Rybicki} \& A.~P. {Lightman}(1986){Rybicki} \& {Lightman}}]{1986rpa..book.....R}
{Rybicki}, G.~B., \& {Lightman}, A.~P. 1986, {Radiative Processes in Astrophysics}

\bibitem[{Y. Sato {et~al.}(2024)Sato, Murase, Bhattacharya, Carpio, Mukhopadhyay, \& Zhang}]{Sato:2024zpq}
Sato, Y., Murase, K., Bhattacharya, M., {et~al.} 2024, \bibinfo{title}{{Two-component off-axis jet model for radio flares of tidal disruption events},} Phys. Rev. D, 110, L061307, \dodoi{10.1103/PhysRevD.110.L061307}

\bibitem[{N. Senno {et~al.}(2017)Senno, Murase, \& Meszaros}]{Senno:2016bso}
Senno, N., Murase, K., \& Meszaros, P. 2017, \bibinfo{title}{{High-energy Neutrino Flares from X-Ray Bright and Dark Tidal Disruption Events},} Astrophys. J., 838, 3, \dodoi{10.3847/1538-4357/aa6344}

\bibitem[{N. Song {et~al.}(2021)Song, Li, Arg{\"u}elles, Bustamante, \& Vincent}]{Song:2020nfh}
Song, N., Li, S.~W., Arg{\"u}elles, C.~A., Bustamante, M., \& Vincent, A.~C. 2021, \bibinfo{title}{{The Future of High-Energy Astrophysical Neutrino Flavor Measurements},} JCAP, 04, 054, \dodoi{10.1088/1475-7516/2021/04/054}

\bibitem[{R. Stein {et~al.}(2021)Stein {et~al.}}]{Stein:2020xhk}
Stein, R., {et~al.} 2021, \bibinfo{title}{{A tidal disruption event coincident with a high-energy neutrino},} Nature Astron., 5, 510, \dodoi{10.1038/s41550-020-01295-8}

\bibitem[{E. Steinberg \& B.~D. Metzger(2018)Steinberg \& Metzger}]{Steinberg:2018pas}
Steinberg, E., \& Metzger, B.~D. 2018, \bibinfo{title}{{The multidimensional structure of radiative shocks: suppressed thermal X-rays and relativistic ion acceleration},} Mon. Not. Roy. Astron. Soc., 479, 687, \dodoi{10.1093/mnras/sty1641}

\bibitem[{N.~C. Stone {et~al.}(2019)Stone, Kesden, Cheng, \& van Velzen}]{Stone:2018nbx}
Stone, N.~C., Kesden, M., Cheng, R.~M., \& van Velzen, S. 2019, \bibinfo{title}{{Stellar tidal disruption events in general relativity},} Gen. Rel. Grav., 51, 30, \dodoi{10.1007/s10714-019-2510-9}

\bibitem[{H. Sun {et~al.}(2015)Sun, Zhang, \& Li}]{Sun:2015bda}
Sun, H., Zhang, B., \& Li, Z. 2015, \bibinfo{title}{{Extragalactic High-energy Transients: Event Rate Densities and Luminosity Functions},} Astrophys. J., 812, 33, \dodoi{10.1088/0004-637X/812/1/33}

\bibitem[{S. {Thorp} {et~al.}(2019){Thorp}, {Chadwick}, \& {Sesana}}]{2019MNRAS.488.4042T}
{Thorp}, S., {Chadwick}, E., \& {Sesana}, A. 2019, \bibinfo{title}{{Tidal disruption events from massive black hole binaries: predictions for ongoing and future surveys},} \mnras, 488, 4042, \dodoi{10.1093/mnras/stz1970}

\bibitem[{S. van Velzen {et~al.}(2024)van Velzen {et~al.}}]{vanVelzen:2021zsm}
van Velzen, S., {et~al.} 2024, \bibinfo{title}{{Establishing accretion flares from supermassive black holes as a source of high-energy neutrinos},} Mon. Not. Roy. Astron. Soc., 529, 2559, \dodoi{10.1093/mnras/stae610}

\bibitem[{M. Wang {et~al.}(2026)Wang, Panagiotou, De, Kara, Masterson, \& Oikonomou}]{Wang:2026ijc}
Wang, M., Panagiotou, C., De, K., {et~al.} 2026, \bibinfo{title}{{Testing the Association of Supermassive Black Hole Infrared Flares and High-energy Neutrinos},} Astrophys. J. Lett., 998, L29, \dodoi{10.3847/2041-8213/ae3f90}

\bibitem[{W. Winter \& C. Lunardini(2021)Winter \& Lunardini}]{Winter:2020ptf}
Winter, W., \& Lunardini, C. 2021, \bibinfo{title}{{A concordance scenario for the observed neutrino from a tidal disruption event},} Nature Astron., 5, 472, \dodoi{10.1038/s41550-021-01343-x}

\bibitem[{W. Winter \& C. Lunardini(2023)Winter \& Lunardini}]{Winter:2022fpf}
Winter, W., \& Lunardini, C. 2023, \bibinfo{title}{{Interpretation of the Observed Neutrino Emission from Three Tidal Disruption Events},} Astrophys. J., 948, 42, \dodoi{10.3847/1538-4357/acbe9e}

\bibitem[{H.-J. Wu {et~al.}(2022)Wu, Mou, Wang, Wang, \& Li}]{Wu:2021vaw}
Wu, H.-J., Mou, G., Wang, K., Wang, W., \& Li, Z. 2022, \bibinfo{title}{{Could TDE outflows produce the PeV neutrino events?},} Mon. Not. Roy. Astron. Soc., 514, 4406, \dodoi{10.1093/mnras/stac1621}

\bibitem[{S.~C. Wu {et~al.}(2025)Wu, Tsuna, Mockler, \& Piro}]{Wu:2025vsr}
Wu, S.~C., Tsuna, D., Mockler, B., \& Piro, A.~L. 2025, \bibinfo{title}{{Delayed radio emission in tidal disruption events from collisions of outflows driven by disk instabilities},} \doarXiv{2511.14008}

\bibitem[{Y. Yao {et~al.}(2023)Yao {et~al.}}]{Yao:2023rbr}
Yao, Y., {et~al.} 2023, \bibinfo{title}{{Tidal Disruption Event Demographics with the Zwicky Transient Facility: Volumetric Rates, Luminosity Function, and Implications for the Local Black Hole Mass Function},} Astrophys. J. Lett., 955, L6, \dodoi{10.3847/2041-8213/acf216}

\bibitem[{Z.~P. Ye {et~al.}(2023)Ye {et~al.}}]{TRIDENT:2022hql}
Ye, Z.~P., {et~al.} 2023, \bibinfo{title}{{A multi-cubic-kilometre neutrino telescope in the western Pacific Ocean},} Nature Astron., 7, 1497, \dodoi{10.1038/s41550-023-02087-6}

\bibitem[{C. Yuan {et~al.}(2024)Yuan, Winter, \& Lunardini}]{Yuan:2024foi}
Yuan, C., Winter, W., \& Lunardini, C. 2024, \bibinfo{title}{{AT2021lwx: Another Neutrino-coincident Tidal Disruption Event with a Strong Dust Echo?},} Astrophys. J., 969, 136, \dodoi{10.3847/1538-4357/ad50a9}

\bibitem[{H. Yuksel {et~al.}(2008)Yuksel, Kistler, Beacom, \& Hopkins}]{Yuksel:2008cu}
Yuksel, H., Kistler, M.~D., Beacom, J.~F., \& Hopkins, A.~M. 2008, \bibinfo{title}{{Revealing the High-Redshift Star Formation Rate with Gamma-Ray Bursts},} Astrophys. J. Lett., 683, L5, \dodoi{10.1086/591449}

\bibitem[{B.~A. Zauderer {et~al.}(2011)Zauderer {et~al.}}]{Zauderer:2011mf}
Zauderer, B.~A., {et~al.} 2011, \bibinfo{title}{{The Birth of a Relativistic Outflow in the Unusual {{\textbackslash}gamma}-ray Transient Swift J164449.3+573451},} Nature, 476, 425, \dodoi{10.1038/nature10366}

\bibitem[{H. Zhang {et~al.}(2025)Zhang, Cui, Huang, Lin, Liu, Qiu, Shao, Shi, Xie, \& Yang}]{Zhang:2024slv}
Zhang, H., Cui, Y., Huang, Y., {et~al.} 2025, \bibinfo{title}{{A proposed deep sea Neutrino Observatory in the Nanhai},} Astropart. Phys., 171, 103123, \dodoi{10.1016/j.astropartphys.2025.103123}

\bibitem[{J.-H. Zheng {et~al.}(2023)Zheng, Liu, \& Wang}]{Zheng:2022afy}
Zheng, J.-H., Liu, R.-Y., \& Wang, X.-Y. 2023, \bibinfo{title}{{Choked Jets in Expanding Envelope as the Origin of the Neutrino Emission Associated with Tidal Disruption Events},} Astrophys. J., 954, 17, \dodoi{10.3847/1538-4357/ace71c}

\end{thebibliography}
\bibliographystyle{aasjournalv7}
\end{document}